\documentclass[11pt]{article}
\usepackage{setspace}
\usepackage{graphicx}
\usepackage[space]{grffile}
\usepackage{amsmath}
\usepackage{amsthm}
\usepackage[sort,comma,authoryear]{natbib}
\usepackage{mathtools}
\usepackage{bm}
\usepackage{amssymb}
\usepackage{threeparttable}
\DeclarePairedDelimiter{\abs}{\lvert}{\rvert}
\usepackage{booktabs}
\usepackage{array}
\usepackage[english]{babel}
\usepackage{longtable}
\usepackage{subfig}
\usepackage{url}
\usepackage{ragged2e}
\usepackage{tabularx}
\usepackage{epstopdf}
\usepackage[toc,title,page]{appendix}
\usepackage{listings}
\newtheorem{theorem}{Theorem}[section]
\newtheorem{lemma}[theorem]{Lemma}
\newtheorem{proposition}[theorem]{Proposition}

\theoremstyle{definition}
\newtheorem{definition}{Definition}[section]

\theoremstyle{remark}
\newtheorem{remark}{Remark}
\usepackage{tabularx}
\usepackage{changepage}
\usepackage{verbatim}
\usepackage[a4paper, total={6in, 8in}]{geometry}
\usepackage{setspace}
\usepackage{booktabs}
\usepackage{geometry}
\usepackage{multirow}
\numberwithin{equation}{section}

\newcommand{\tabincell}[2]{\begin{tabular}{@{}#1@{}}#2\end{tabular}}

\usepackage{hyperref}
\usepackage{natbib}
\hypersetup{
	colorlinks,
	linkcolor={blue},
	citecolor={blue},
	urlcolor={blue}
}

\begin{document}
	\author{  
	Zhengxiao Li \footnote{Corresponding author: School of Insurance, University of International Business and Economics, Beijing, China. Email: li\_zhengxiao@uibe.edu.cn.}
		\quad
		Jan Beirlant \footnote{Dept. of Mathematics, LStat and LRisk, KU Leuven, Belgium, and Dept. of Mathematical Statistics and Actuarial Science, Univ. of the Free State, South Africa.}
		\quad Liang Yang \footnote{School of Insurance, Southwestern University of Finance and Economics, Chengdu, China.} 
	}
	\date{}
	\title{A new class of copula regression models for modelling multivariate heavy-tailed data}

	\maketitle
	\begin{abstract}
A new class of copulas, termed the  MGL copula class, is introduced. The new copula originates
from extracting the dependence function of the multivariate generalized log-Moyal-gamma  distribution
whose marginals follow the univariate generalized log-Moyal-gamma (GLMGA) distribution as introduced in \citet{li2019jan}. The MGL
copula can capture nonelliptical, exchangeable, and asymmetric dependencies among marginal coordinates and provides a simple formulation for regression applications. 
We discuss
the probabilistic characteristics of MGL copula and obtain the corresponding extreme-value
copula, named the MGL-EV copula. 
While the survival MGL copula can be also regarded as a special case of the MGB2 copula from \citet{yang2011generalized}, 
we show that the proposed model is effective in regression modelling of dependence structures.
Next to a simulation study, we propose  
two applications illustrating the usefulness of the
proposed model. 
This method is also implemented in a user-friendly R package: \texttt{rMGLReg}.
		\\
		\\
		\\
		{\bf{Article History:}} Compiled \today
	\end{abstract}
	\newpage

\section{Introduction}\label{section:introduction}
Modelling multivariate heavy-tailed data  is an important challenge in actuarial statistics.
While univariate  risk models based on heavy tailed distributions are well developed (see e.g. \citet{beirlant2003regression}, \cite{li2016bayesian}, \citet{Leppisaari2016Modeling}, and \citet{li2019jan}), predicting extreme loss through multivariate models has received much less attention, especially in the presence of additional covariate information.

Copulas have become highly popular in modelling flexible dependence structures of multi-dimensional data during the past several decades.
The advantage of the copula concept is that it separates  modelling of the marginals from the dependence structure \citep{joe2014dependence}.
Multivariate loss often exhibits asymmetric dependence, with special emphasis on the joint upper tail dependence of the multivariate heavy tailed data.
Elliptical copulas such as the Gaussian or Student-t copulas, and Archimedean copulas with
exchangeable dependence structures,  imply symmetric dependence.
Extreme-value copulas are tailored for extremes and can capture tail dependence and asymmetry, while computation of joint densities can be excessively prohibitive in higher dimensions \citep{castruccio2016high}.
Hence, copula families that are able to accommodate tail dependence and to
capture dependence asymmetry,
 and are easy to implement, are highly desirable.

Some other methods have also been proposed to construct new classes of copulas in the literature, such as the
Laplace transform method \citep{yang2020family},
the geometric weighting method \citep{zhang2016new},
the vine copula approach \citep{aas2009pair,shi2018pair},  and
the factor copula model \citep{oh2017modeling}, among others.
In order to model multivariate heavy-tailed data,
\cite{yang2011generalized} extracted a new copula family  from the multivariate generalized beta distribution of the second kind, named the generalized beta copula or MGB2 copula, which features positive tail dependence in the joint upper tail and tail independence in the joint lower tail.

The main contribution of this article is to propose a new class of beta-type copulas, the MGL copula family, where 	the dependence function is extracted from a new multi-dimensional version
of the univariate GLMGA distribution which was proposed in \cite{li2019jan}.
We provide some important characteristics of this class
and obtain the corresponding extreme-value copula (the MGL-EV copula).
Inheriting heavy-tail features from the GLMGA distribution and its multivariate
extension, 
the  MGL and MGL-EV copulas
are able to account for joint
extreme events based on a positive tail dependence index.
Furthermore, these new copulas can accommodate non-elliptical and
asymmetric dependence, and are easily extended for regression applications.
We illustrate the usefulness of the proposed copula regression modelling using two insurance cases.
The first example is about modelling the pairwise dependence between two continuous variables in the well-known Danish fire insurance data set, which was already considered for instance   in \cite{hashorva2017some} and \cite{lu2021nonparametric}.
In the second application  we  model the dependence between a continuous and a semi-continuous variable in a Chinese earthquake data set. This data set was already analyzed in \cite{li2019jan} concerning the  univariate losses. In both case studies we consider the evolution of the dependence as a function of time.
We compare the performance of the proposed  models with the MGB2 copula and  other copula candidates  in terms of goodness of fit and  tail dependence measures.

The remainder of the paper is structured as follows.
In Section \ref{section:MGL copula}, after recapitulating the univariate GLMGA distribution, we construct a  multivariate extension and the  corresponding copula.
In Section \ref{section:properties} we report properties of the proposed copulas 
and obtain the corresponding extreme-value
copulas.
The copula regression modelling is discussed in Section \ref{section: regression}.
A simulation study is conducted in Section \ref{section: simulation} and the two  applications are discussed  in Section \ref{section:application}.
Finally we formulate some conclusions and future possible extensions. Proofs are deferred to the Appendix.
The R package: \texttt{rMGLReg} can be found at \url{https://github.com/lizhengxiao/rMGLReg}.


	\section{From a multivariate GLMGA distribution to a MGL copula}\label{section:MGL copula}
		The GLMGA three-parameter distribution model as proposed in \cite{li2019jan}, is obtained by mixing a generalized log-Moyal distribution (GlogM)   \citep{bhati2018generalized} with the gamma distribution.
		\begin{definition}
		The random variable $Y>0$ follows a GLMGA distribution ($Y\sim \text{GLMGA}(\sigma,a,b)$) if it admits the following stochastic representation:
		\[
		Y|\Theta \sim \text{GlogM}(\Theta,\sigma)\quad \text{and}\quad \Theta \sim \text{Gamma}(a, b),
		\]
		where $\text{GlogM}(\Theta,\sigma)$ refers to the generalized log-Moyal distribution introduced in \cite{bhati2018generalized} with density and distribution function (cdf)
		\begin{align}
		{f}_{Y|\theta}(y|\theta)&=\frac{\sqrt{\theta}}{\sqrt{2\pi }\sigma }\ {{\left( \frac{1}{y} \right)}^{\frac{1}{2\sigma }+1}}{{\exp}{\left[-\frac{\theta}{2}{{\left( \frac{1}{y} \right)}^{1/\sigma}}\right]}},\nonumber \\
		{F}_{Y|\theta}(y|\theta)&=\text{erfc}\left(\sqrt{{\theta}/{2}}y^{-\frac{1}{2\sigma}}\right).
		\label{eq:cdf-GlogM}
		\end{align}
Here $\text{Gamma}(a, b)$ refers to the gamma distribution with density
		\begin{equation}\label{pdf:gamma}
		p(\theta) =\frac{{b }^{a }}{\Gamma \left( a\right)}\theta^{a-1}\exp{\left(-b \theta\right)},
		\end{equation}
		for $a,b>0$, while $\text{erfc}(\cdot)$ denotes the complementary error function given by $\text{erfc}(z)=\frac{2}{\sqrt{\pi}}\int_{z}^{\infty}\exp(-t^2)dt$.
		\end{definition}
		The density function of the GLMGA distribution is then given by 
		\begin{align}
		f\left(y \right)=\frac{(2b)^a}{\sigma B(a,\frac{1}{2})}\frac{y^{-\left(\frac{1}{2\sigma}+1\right)}}{\left(y^{-\frac{1}{\sigma}}+2b\right)^{a + \frac{1}{2}}},
		\label{pdf:LMGA}
		\end{align}
		for $y>0$, $\sigma>0, a>0, b>0$, with  $B(m,n)=\int_0^{1}t^{m-1}(1-t)^{n-1}dt$ the beta function.
	The cdf and quantile function of
	the GLMGA distribution are  given by
	\begin{align}
	\label{cdf:GLMA}
	F\left(y\right)&=1-I_{\frac{1}{2},a}\left(\frac{y^{-1/\sigma}}{y^{-1/\sigma}+2b}\right), \; y>0,\\
	F^{-1}(p)&=(2b)^{-\sigma}\left[\frac{I^{-1}_{\frac{1}{2},a}(1-p)}{1-I^{-1}_{\frac{1}{2},a}(1-p)}\right]^{-\sigma},
	\; p \in (0,1),
	\label{qf:GLMA}
	\end{align}
	where  $I^{-1}_{m,n}(\cdot)$ denotes the inverse of the beta cumulative distribution function $I_{m,n}(\cdot)$  (or regularized incomplete beta function).

	Moreover, the survival function $\bar{F}=1-F$ of the GLMGA distribution allows for the expansion at infinity
	\begin{equation}
		\bar{F}(y)= Cy^{-1/(2\sigma)}\left\{1 + Dy^{-1/\sigma}\left(1+o(1)\right)\right\}, \quad \quad y\to +\infty,
		\label{Patype}
		\end{equation}
		with $C=\frac{2}{(2b)^{1/2}B(a,\frac{1}{2})}>0$ and $D=-\frac{2a+1}{12b}$. Hence, the GLMGA distribution is of Pareto-type with Pareto tail index $1/(2\sigma)$.
		
		Also, near 0 the distribution function $F$ is regularly varying with index 
		$a/\sigma$:
		\begin{equation}
		\lim_{t \to 0+} \frac{F(ty)}{F(t)}= y^{a/\sigma}, \mbox{ for all } y>0. 
		\label{RV0}
		\end{equation}
		
		Finally, we mention that the $\text{GLMGA}(\sigma,a,b)$ distribution is a special case of the four-parameter generalized beta distribution of the second kind $\text{GB2}(\tau, \mu, \nu, p)$ by substituting  $\tau=a,\mu=(2b)^{-\sigma},\nu=\frac{1}{2}$ and $p=-\frac{1}{\sigma}$ with the density function given by
		$
		f_{\text{GB2}}\left(y \right)=\frac{\abs{p}}{B(\nu,\tau)y}\frac{\mu^{p \tau}y^{p\nu}}{(y^p+\mu^p)^{\nu+\tau}}
		$.	\\
		
	As noted in \cite{li2019jan}, the  univariate GLMGA distribution can be  used to accommodate the extreme risks and capture both tail and modal parts of heavy-tailed insurance data, and it occupies an interesting position in between the popular GB2 model and its subfamilies, such as the Lomax model.

	\subsection{The multivariate GLMGA distribution (MGL)}\label{section:MGL distribution}
	
	In this section we propose a multivariate extension of  the univariate GLMGA distribution as a gamma mixture of the GlogM distribution from \cite{bhati2018generalized} using a common scale parameter $a$  over all dimensions.
	
	\begin{definition}
		A $d$-dimensional random vector $\bm{Y}=(Y_{1},\dots,Y_{d})^T$  on $(0,\infty )^d$ follows a multivariate GLMGA distribution (denoted by $\bm{Y} \sim \text{MGL}(\bm{\sigma}, a, \bm{b})$),
		with $\bm{\sigma} = (\sigma_1,\ldots, \sigma_d)$ and $\bm{b} = (b_1,\ldots, b_d)$ if 
		\begin{itemize}
		\item each $Y_j$ given $\Theta = \theta$ follows a $\text{GlogM}(\theta/b_j,\sigma_j)$
		with  density
		\begin{equation}\label{pdf:cGLMGA}
		f_{Y_j|\Theta = \theta}(y_j|\theta)
		=\frac{\sqrt{\theta/b_j}}{\sqrt{2\pi }\sigma_j }\ {{\left( \frac{1}{y_{j}} \right)}^{\frac{1}{2\sigma_j }+1}}{{\exp}{\left[-\frac{\theta/b_j}{2}{{\left( \frac{1}{y_{j}} \right)}^{1/\sigma_j}}\right]}}, \quad y_j>0,
		\end{equation}
		where $b_j>0$ and $\sigma_j>0$,
		\item 
	 $Y_{1},\dots,Y_{d}$ are conditionally independent given $\Theta$,
	 \end{itemize}
	 where  the mixing variable $\Theta$ follows a gamma distribution with shape parameter $a$ and one unit rate, i.e. $\Theta\sim \text{Gamma}(a,1)$.
\end{definition}	 
	 
		The above definition easily leads to the following multivariate GLMGA density by taking the expectation with respect to $\Theta$:
		\begin{align}
		f(y_1,\dots,y_d)&
		=\frac{\Gamma(a+\frac{d}{2})}{\Gamma(a)\Gamma(\frac{1}{2})^d\prod_{j=1}^{d}\sigma_j y_j}
		\frac{\prod_{j=1}^{d}\left[(2b_j)^{\sigma_j}y_j\right]^{-\frac{1}{2\sigma_j}}}
		{\left[\sum_{j=1}^{d}\left((2b_j)^{\sigma_j}y_j\right)^{-\frac{1}{\sigma_j}}+1\right]^{a+\frac{d}{2}}}
		\label{pdf:MLMGA},
		\end{align}
		for $y_{j} >0$, being $\sigma_j>0, a>0, b_j>0$.\\
		Since $Y_j$ ($j=1,\ldots,d$) are conditionally independent given $\Theta$, the marginal distributions are obtained  by setting $d=1$ which leads to the densities in \eqref{pdf:LMGA}.\\

Moments of the MGL distribution are easy to calculate thanks to the gamma mixture structure.
	\begin{proposition}		\label{prop:corelation}
		Suppose $\bm{Y} \sim \text{MGL}(\bm{\sigma}, a, \bm{b})$.  Then, when $\max_j \sigma_j < 1/4$
		\begin{align}
		\mathbb{E}(\bm{Y})&=\mathbb{E}\left[\mathbb{E}(\bm{Y}|\theta)\right]=
		(2b_j)^{-\sum_{j=1}^d\sigma_j}\frac{\prod_{j=1}^{d}\Gamma(\frac{1}{2}-\sigma_j)}{\Gamma(\frac{1}{2}-\sum_{j=1}^d\sigma_{j})}
		\frac{B(\frac{1}{2}-\sum_{j=1}^{d}\sigma_j, \sum_{j=1}^d\sigma_{j}+a)}{B(\frac{1}{2},a)},
		\\
		\text{Cov}(\bm{Y})&=\mathbb{E}\left[\text{Var}\left(\bm{Y}|\theta \right)\right]
		+ \text{VaR}\left[\mathbb{E}\left(\bm{Y}|\theta \right)\right]
		=\bm{\Sigma},
		\end{align}
		where the components $\Sigma_{jj'} = \text{Cov}(Y_j,Y_{j'})$ of the variance-covariance matrix $\bm{\Sigma}$ are given by 
\[
		\Sigma_{jj'} = \mathbb{E}(Y_j)\mathbb{E}(Y_{j'})\left[
		\frac{B(a+\sigma_j+\sigma_{j'},a)}{B(a+\sigma_j,a+\sigma_{j'})}
		-1\right], \; j\neq j'. \]
The  correlations  are given by
		\begin{equation}\label{pairwise correlation}
		\text{Corr} (Y_j,Y_{j'})
		=\rho_{jj'}
		=
		\frac{\frac{B(a+\sigma_j+\sigma_{j'},a)}{B(a+\sigma_j,a+\sigma_{j'})}
			-1}{
			\sqrt{\left[\frac{B(a+2\sigma_j,a)B(\frac{1}{2}-2\sigma_j,\frac{1}{2})}{B(a+\sigma_j,a+\sigma_j)B(\frac{1}{2}-\sigma_j,\frac{1}{2}-\sigma_j)}
				-1\right]\left[\frac{B(a+2\sigma_{j'},a)B(\frac{1}{2}-2\sigma_{j'},\frac{1}{2})}{B(a+\sigma_{j'},a+\sigma_{j'})B(\frac{1}{2}-\sigma_{j'},\frac{1}{2}-\sigma_{j'})}
				-1\right]}},
	\quad j\ne j'.	
	\end{equation}
	Moreover 
	\[
	\lim_{a \to \infty} \rho_{jj'} =0,  \mbox{ for } j \neq j'.
	\]
	\end{proposition}
	
	
Next we show that the MGL distribution is closed under conditional distributions. This result can be used in order to simulate random samples from the MGL distribution.
This property also allows insurers to  
derive 
the conditional mean, $\mathbb{E}(Y_{r+1}|Y_1,...,Y_r)$,
 incorporating past experience claim amount $Y_1,...,Y_r$ into future premium in a nonlinear fashion
for experience ratemaking application in non-life actuarial science  \citep{shi2018pair}.

	\begin{proposition}\label{prop: conditional}
	Suppose $\bm{Y} \sim \text{MGL}(\bm{\sigma}, a, \bm{b})$ and consider two complementary sub-vectors
	$\bm{Y_1}=(Y_1,\cdots,Y_r)^T$ and $\bm{Y_2}=(Y_{r+1},\cdots, Y_d)^T$   of $\bm{Y}$.
	Then the conditional distribution  of $\bm{Y_1}$ given $\bm{Y_2}$ equals $\text{MGL}(\bm{\sigma}, a^*, \bm{b}^*)$
		where $a^{*}=a+\frac{d-r}{2}$ and
		$
		b^{*}_j=b_j \left[1+\sum_{j=r+1}^d y_j^{-1/\sigma_j}/(2b_j)\right]
		$.
	\end{proposition}

\begin{figure}[htbp]
		\includegraphics[scale=0.5]{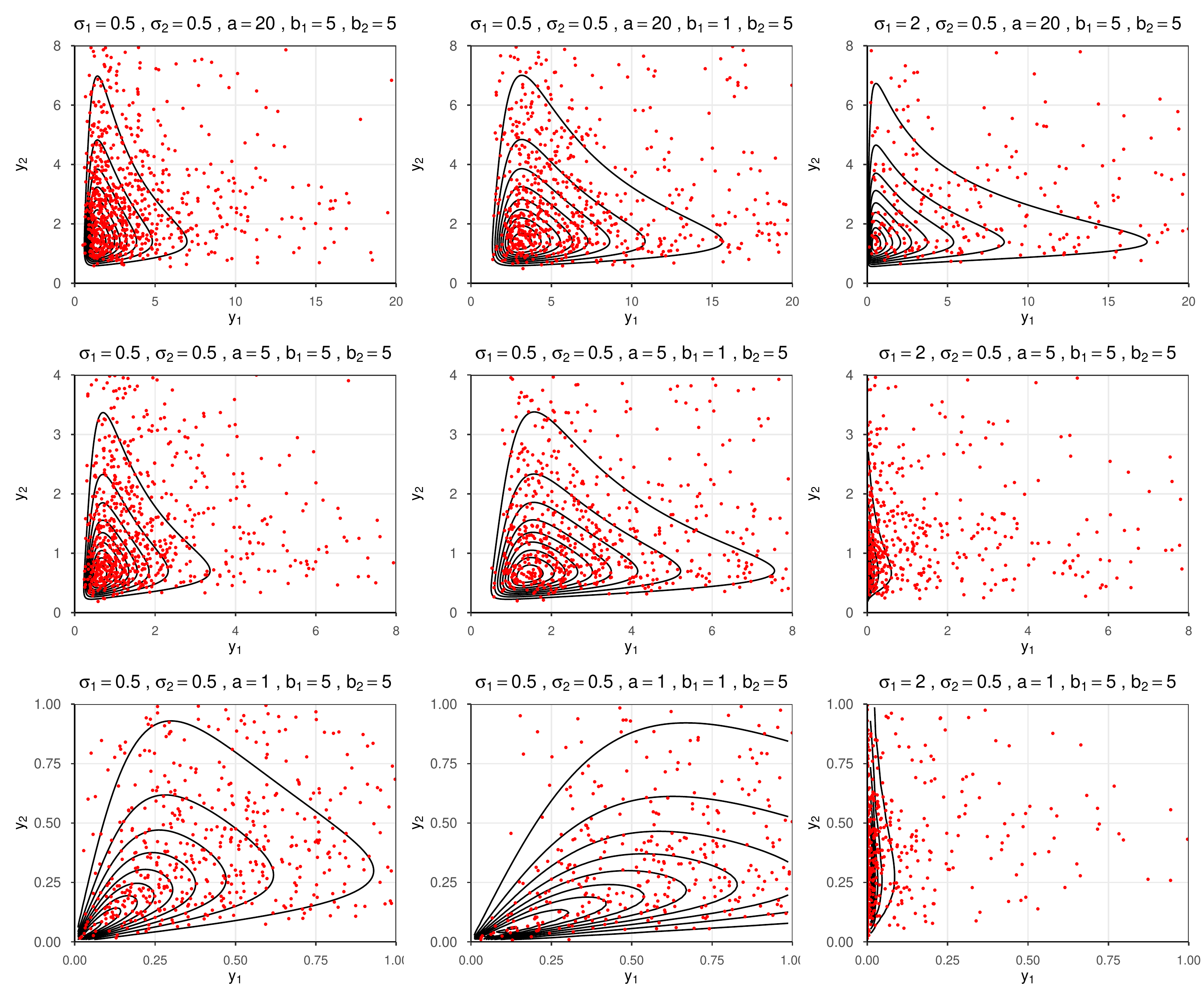}
					\caption{Contour plots of the MGL distribution with the simulated sample 1,000 (red points) when different values of parameters $(\sigma_1, \sigma_2,a,b_1,b_2)$ are considered.
	}
		\label{fig:counterplot-MGL}
	\end{figure}

Figure \ref{fig:counterplot-MGL} displays the scatter for the MGL distribution with simulated sample size 1,000 for different combinations of  $(\sigma_1,\sigma_2, a, b_1, b_2)$. Contour plots of the density function are also given in Figure \ref{fig:counterplot-MGL}.
	Positive dependences and the tail asymmetry features in the observed MGL data.
	The dependence parameters appear in the marginal distributions and 
	the MGL features stronger lower tail dependence with smaller values of $a$.

%

%
%

	\subsection{MGL copula and survival MGL copula}\label{section: MGL}
	Although the MGL distribution may provide a useful tool for handling the multivariate heavy-tailed
	data, 	it suffers some limitations: 
	the univariate marginal
	distributions belong to the same family, 
	the dependence parameters  appear in the marginal distributions \citep{frees1998understanding, yang2011generalized}, and each margin contains the common parameter $a$.
	Considering the corresponding MGL copula and survival MGL copula allows to separate the modelling of the marginal and dependence	structures.
	Based on Sklar's theorem \citep{sklar1959fonctions}, any joint cdf $H$ with continuous marginal cdfs $F_1...,F_d$ for a sequence of random variables $Y_{1},\dots,Y_{d}$ has  a unique copula $C$ through
	\[
	H(y_1,\dots,y_d)=C(F_1(y_1),\dots,F_d(y_d)),
	\]
	where $C$ represents a multivariate joint distribution defined on a $d$-dimensional cube $(0,1)^d$ such that every marginal follows the uniform (0,1) distribution. It is convenient to rewrite this as 
	\begin{equation*}
	H(y_1,\dots,y_d)=C(u_1,\dots,u_d),
	\end{equation*}
	with $u_j=F_j(y_j)$, $j=1,\dots,d$.

	\begin{definition}
		The MGL copula is defined as 
		\begin{align}
		C^{\text{MGL}}(u_1, \dots, u_d;a) &= H(F_{1}^{-1}(y_1),...,F_{d}^{-1}(y_d)), \quad (u_1, \dots, u_d)\in (0,1)^{d} \nonumber\\
		&=\mathbb{E}_{\Theta}\left[\prod_{j=1}^{d}\text{erfc}\left(
		\sqrt{t(u_j;a)\Theta}
		\right)\right],
		\label{cdf:copula}
		\end{align}
		with $\Theta \sim \text{gamma}(a,1)$ and 
		\[		t(u_j;a)=\frac{I^{-1}_{\frac{1}{2},{a}}(1-u_j)}{1-I^{-1}_{\frac{1}{2},{a}}(1-u_j)}.
		\]
		\end{definition}
		
				The
		corresponding copula density function is given by
\begin{align}
		c^{\text{MGL}}(u_1, \dots, u_d;a)&=
		\frac{h(F_{1}^{-1}(u_1), ..., F_{d}^{-1}(u_d))}{\prod_{j=1}^{d}f_j(F_{j}^{-1}(u_d))} \nonumber\\
		&=\frac{\Gamma(a)^{d-1}\Gamma(a+\frac{d}{2})}{\Gamma(a+\frac{1}{2})^d}\frac{\prod_{j=1}^{d}(t(u_j;a)+1)^{a+\frac{1}{2}}}{\left(\sum_{j=1}^{d}t(u_j;a)+1\right)^{a+\frac{d}{2}}},
		\quad (u_1, ..., u_d)\in \left[0,1\right]^d,
		\label{pdf:copula}
		\end{align}
		where $h$ denotes the joint density of the $\text{MGL}(\bm{\sigma}, a, \bm{b})$ distribution and $f_j$ is the density of the univariate GLMGA distribution with parameters $(\sigma_j, a, b_j)$. \\

Given that larger values of the common parameter $a$ in the  MGL distribution yield weaker dependence, in what follows we re-parameterize the MGL copula by substituting $\delta=1/a$, and the density and cdf are denoted with
		$c^{\text{MGL}}(\cdot;\delta)$ and $C^{\text{MGL}}(\cdot;\delta)$. \\

We next propose a simulation procedure for pseudo data from the MGL copula.

	\begin{proposition}\label{prop: simulation}
Pseudo random vectors  from  $C^{MGL}(\cdot;\delta)$ can be constructed using the following steps:
		\begin{itemize}
			\item Generate i.i.d. random samples $(U_1, ..., U_d)$ from the uniform (0,1) distribution,
			and let $Z_j=\frac{I_{\frac{1}{2}, k_j}^{-1}(1-U_j)}{1-I_{\frac{1}{2}, k_j}^{-1}(1-U_j)}$, where $k_j =\frac{1}{\delta} + \frac{j- 1}{2}$ for $j=1,...,d$;
			\item Generate the random numbers 
			$
			M_1 =Z_1$, and then $M_j=(1+\sum_{k=1}^{j-1}{M_{k}})Z_j$ for $j=2,...,d$;
			\item Compute $U_j^{*}=1-I_{\frac{1}{2},\frac{1}{\delta}}(\frac{M_j}{1 + M_j})$ for $j = 1, ..., d$.
		\end{itemize}
			\end{proposition}

\begin{figure}[htbp]
		\includegraphics[scale=0.55]{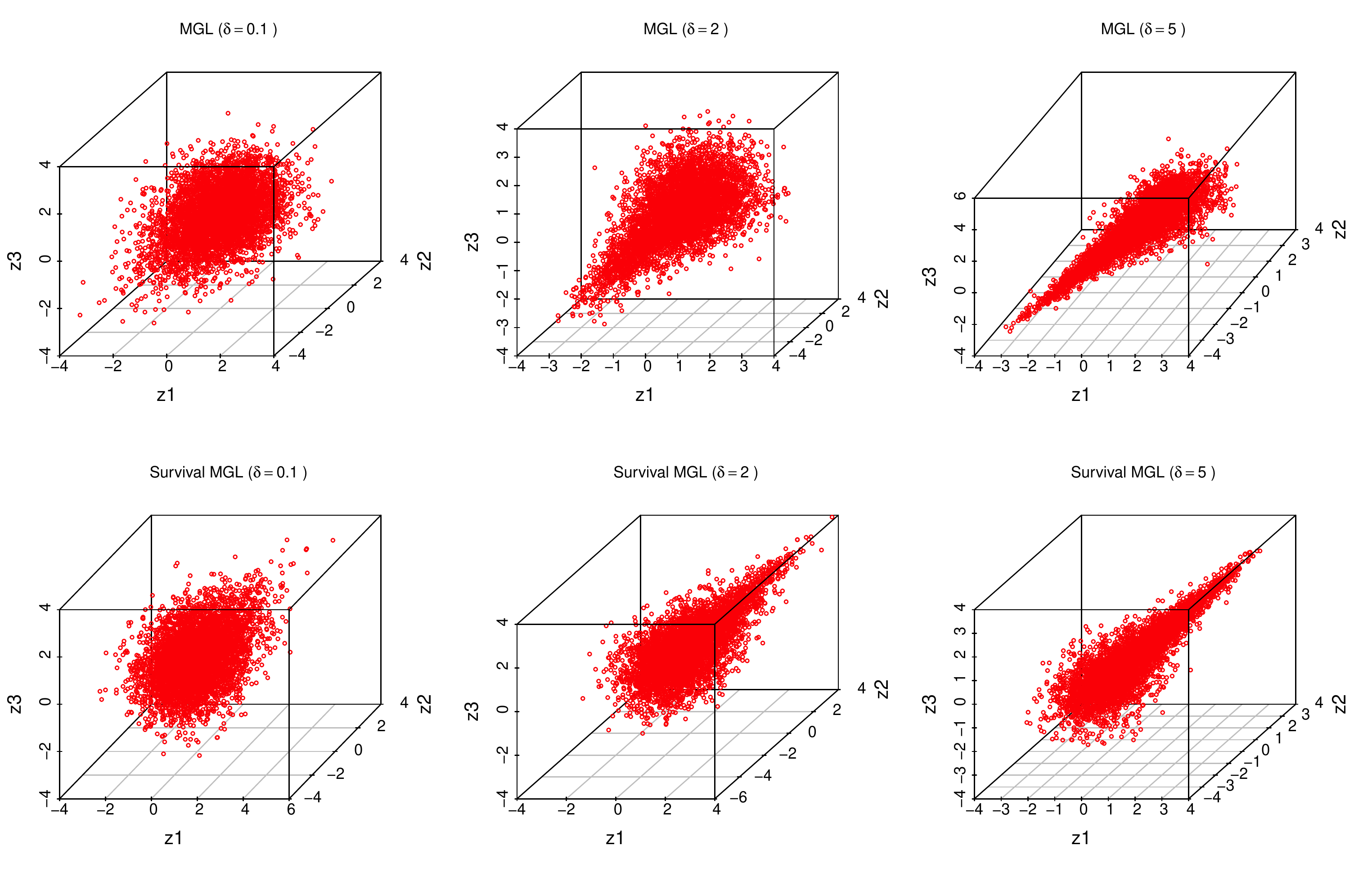} 
		\caption{
			The	normalized scatter plots 
			with simulated sample size 5,000 for low ($\delta=0.1$), medium ($\delta=2$) and high dependence ($\delta=5$) of the 3-dimensional
         MGL copula (top row) and survival MGL copula (bottom row).
		}
		\label{fig:plot3d-copula}
	\end{figure}

	To illustrate the dependence structure, in Figure \ref{fig:plot3d-copula}  simulated normalized scatter plots with $d=3$  from the MGL copula are given with $n=5,000$ for  different values of copula parameter $\delta$. 
	The normalized random samples are defined as $z_{ij}=\Phi^{-1}(u_{ij}^{*})$ for $j=1,2,3$ and $i = 1,...,5,000$, where $\Phi(\cdot)$ is the standard normal distribution function and $u_{ij}^{*}$ denote the copula  realizations. 
	Note the positive dependences  and the tail asymmetry features among the three 
	variables, while  the MGL copula features stronger tail dependence in the lower tail with the larger value of the parameter $\delta$. \\
	
	To capture the upper tail dependence structure, we propose a survival MGL copula $\bar{C}^{MGL}(\cdot;\delta)$.
	\begin{definition}
		The survival MGL copula is defined as 
		\begin{equation}
		\bar{C}^{MGL}(u_1, ..., u_d;\delta) =
		1-\sum_{j=1}^d(1-u_j)+
		\sum_{J\subseteq\left\{1,...,d\right\}}(-1)^{\abs J}C^{MGL}((1-u_1)^{1(1\in J)}, ..., (1-u_d)^{1(d\in J)};\delta), 
		\end{equation}
		where the sum extends over all $2^d$ subsets $J$ of $\left\{1,...,d\right\}$ and  $\abs J$ denotes the number of elements of $J$ and $1(j\in J)$  the indicator of $J$. 
			For $d=2$, 
		$\bar{C}^{MGL}(u_1,u_2;\delta)=u_1+u_2-1+C^{MGL}(1-u_1,1-u_2;\delta)$.	
			\end{definition}

			The density function of the survival MGL copula  is given by
			\begin{equation}
				\bar{c}^{MGL}(u_1, ...,u_d;\delta) = {c}^{MGL}(1-u_1, ...,1-u_d;\delta).
			\end{equation}
			The one-parameter survival MGL with joint density function $\bar{c}^{MGL}(u_1, ...,u_d;\delta)$ can be regarded as a special case of the  MGB2 copula (see \cite{yang2011generalized}) with $(d+1)$-parameters  and joint density function $c^{MGB2}(u_1, ...,u_d;p_1,...p_d,q)$ given by
			\[
			c^{MGB2}(u_1, ...,u_d;p_1,...p_d,q)=\frac{\Gamma(q)^{d-1}\Gamma\left(\sum_{i=1}^d p_i + q\right)}{\prod_{i=1}^d\Gamma(p_i+q)}\frac{\prod_{i=1}^d \left(1 + x(u_i;p_i,q) \right)^{p_i+q}}{\left(1 + \sum_{i=1}^{d}x(u_i;p_i,q)\right)^{\sum_{i=1}^{d}p_i+q}},
			\]
			in which $x(u_i;p_i,q) = I_{p_{i},q}^{-1}(u_i)/\left(1 - I_{p_{i},q}^{-1}(u_i)\right)$. The survival MGL density is obtained taking
				 $p_1= p_2= \ldots = p_d=\frac{1}{2}$ and $q=\delta$.

	\section{Properties of the MGL copula}\label{section:properties}

	In this section we collect some basic properties of the MGL copula and specify these in the bivariate case.
	The $h$-function represents the conditional distribution function of a bivariate copula, which is defined as the partial derivative of the distribution function of the copula with respect to the first argument $h_{2|1}(u_2|u_1):=\Pr(U_2\le u_2|U_1= u_1)=\partial C(u_1,u_2)/\partial u_1$, and with respect to second argument $h_{1|2}(u_1|u_2):=\Pr(U_1\le u_1|U_2= u_2)=\partial C(u_1,u_2)/\partial u_2$ \citep{schepsmeier2014derivatives}.
	\begin{proposition}		
		The $h$-function corresponding to the bivariate MGL copula is given by 
		\begin{align}
		h_{2|1}^{\text{MGL}}(u_2|u_1;\delta)&= 1-I_{\frac{1}{2},\frac{1}{\delta}+\frac{1}{2}}\left[
		\frac{t(u_2;\delta)}{t(u_1;\delta) + t(u_2;\delta)+1}
		\right],
		\label{h1-function}\\
		h_{1|2}^{\text{MGL}}(u_1|u_2;\delta) &= 1-I_{\frac{1}{2},\frac{1}{\delta}+\frac{1}{2}}\left[
		\frac{t(u_1;\delta)}{t(u_1;\delta) + t(u_2;\delta)+1}
		\right],
		\label{h2-function}
		\end{align}
for all $(u_1, u_2)\in\left[0,1\right]^2$.		
	\end{proposition}
	\noindent
This result follows combining Proposition \ref{prop: conditional}, 	\ref{cdf:GLMA} and \ref{qf:GLMA}.
	
	\begin{remark}
		Bivariate random samples can be generated using the inverse conditional distribution function of a given parametric bivariate copula.
		This function can be easily inverted using the quantile function
		of the beta distribution, yielding the corresponding inverse $h$-function of MGL copula
		\begin{align}
		h_{2|1}^{-1}(u_2|u_1;\delta)&=1-I_{\frac{1}{2},\frac{1}{\delta}}\left[
		\frac{(t(u_1;\delta)  + 1)t(u_2;2\delta/(2+\delta))}
		{(t(u_1;\delta)  + 1)t(u_2;2\delta/(2+\delta)) + 1}
		\right],
		\label{inverse h1-function}\\
		h_{1|2}^{-1}(u_1|u_2, {\delta})&=1-I_{\frac{1}{2},\frac{1}{\delta}}\left[
		\frac{(t(u_2;\delta)  + 1)t(u_1;2\delta/(2+\delta))}
		{(t(u_1;\delta)  + 1)t(u_2;2\delta/(2+\delta)) + 1}
		\right].
		\label{inverse h2-function}
		\end{align}
	\end{remark}
	
\noindent	
{\bf Rank based measures of association}
	Kendall's \textit{tau} and Spearman's \textit{rho} are two well-known rank
	based measures of association in copula modelling. Unlike Pearson correlation coefficient,
	Kendall's \textit{tau} and Spearman's \textit{rho} solely depend on the copula
	function $C$  through
	\begin{align*}
	\tau&=4\int_{0}^{1}\int_{0}^{1}C(u,v)dC(u,v)-1,\\
	\rho&=12\int_{0}^{1}\int_{0}^{1}C(u,v)dudv-3.
	\end{align*}
	\citep{nelsen2007introduction, fredricks2007relationship}
	\\
	For a bivariate copula these correlations can be evaluated using
	\begin{align*}
	\tau &= 1-4\int_{\left[0,1\right]^2}h_{2|1}(v|u)h_{1|2}(u|v)dudv,
	\\
	\rho &=3-12\int_{\left[0,1\right]^2}h_{2|1}(v|u)ududv,
	\end{align*}
	where $h_{2|1}(\cdot|\cdot)$ and $h_{1|2}(\cdot|\cdot)$ denote the $h$-function of the bivariate MGL copula given in (\refeq{h1-function}) and (\refeq{h2-function}) respectively.
	Although there is no closed form solution,
	the integrations on the support of unit hypercubes can well be
	approximated using numerical methods. \\

	\noindent
{\bf Tail dependence behaviour}
	It is important to understand how the parameter $\delta\in(0, \infty)$ influences the level of
	dependence in the the bivariate MGL copula. One can show that the copula approaches the independence copula as $\delta$
	approaches 0, that is
	\begin{equation*}\label{ind copula}
	\lim_{\delta\to +0}C^{MGL}(u_1,u_2;\delta)=u_1u_2.
	\end{equation*}
	Moreover, it is easy to check that the copula
	becomes singular as $\delta \to +\infty$. \\

	Tail dependence, also known as extremal dependence or
	asymptotic dependence, quantifies the probability of concurrence
	of extreme events in the upper tail or lower tail of a bivariate distribution. 
The indices of lower and upper tail dependence
	of a copula $C$ are defined by
	\begin{align*}
	&\lambda_{l}= \lim_{ u \to 0^+} \mathbb{P}(U_2 \leq u | U_1 \leq u)=\lim_{ u \to 0^+}\frac{ C(u,u)}{ u}, \\
	&\lambda_{u}=\lim_{ u \to 1^-} \mathbb{P}(U_2 > u | U_1 > u)=\lim_{u \to 1^-}\frac{1-2u+C(u,u)}{ u},
	\end{align*}
	provided the limits $\lambda_l$ and $\lambda_{u}$ exist in $\left[0,1\right]$ \citep{joe1997multivariate}.
	The copula $C$ is said to be
	lower, respectively  upper,  asymptotically tail dependent if $\lambda_l\ne 0$, respectively $\lambda_u\ne 0$.
	Moreover, copulas of elliptically symmetric distributions have $\lambda_l=\lambda_u$.

\begin{figure}[htbp]
		\centering
		\includegraphics[scale=0.35]{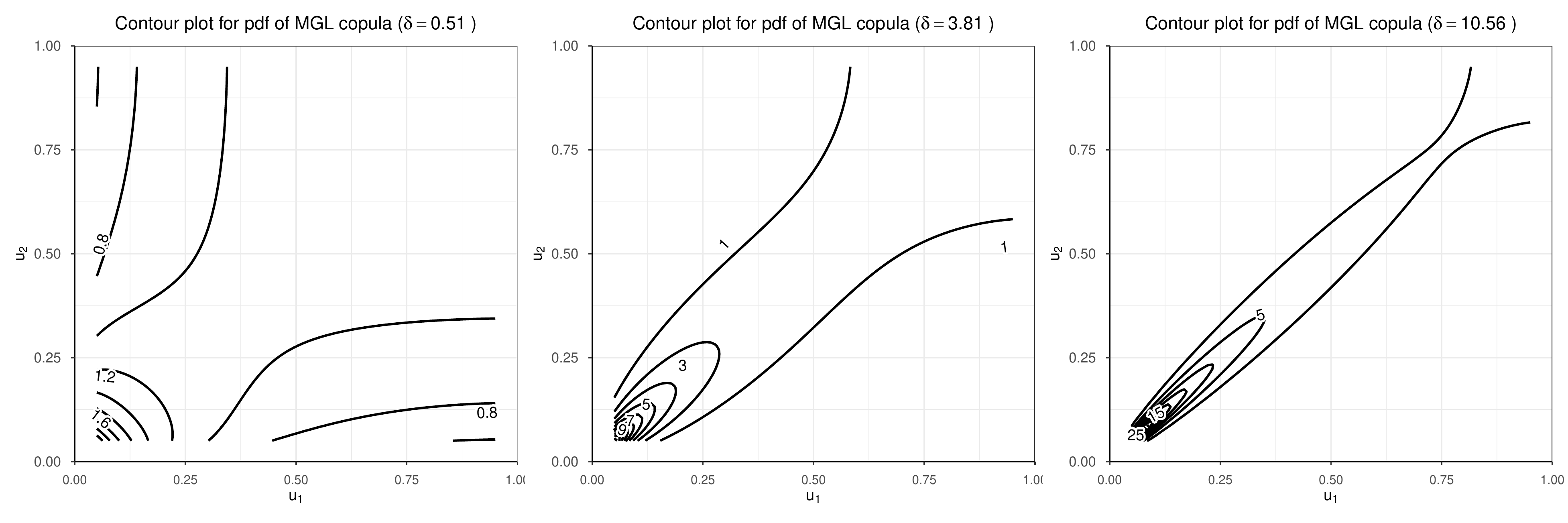} \\
		\includegraphics[scale=0.35]{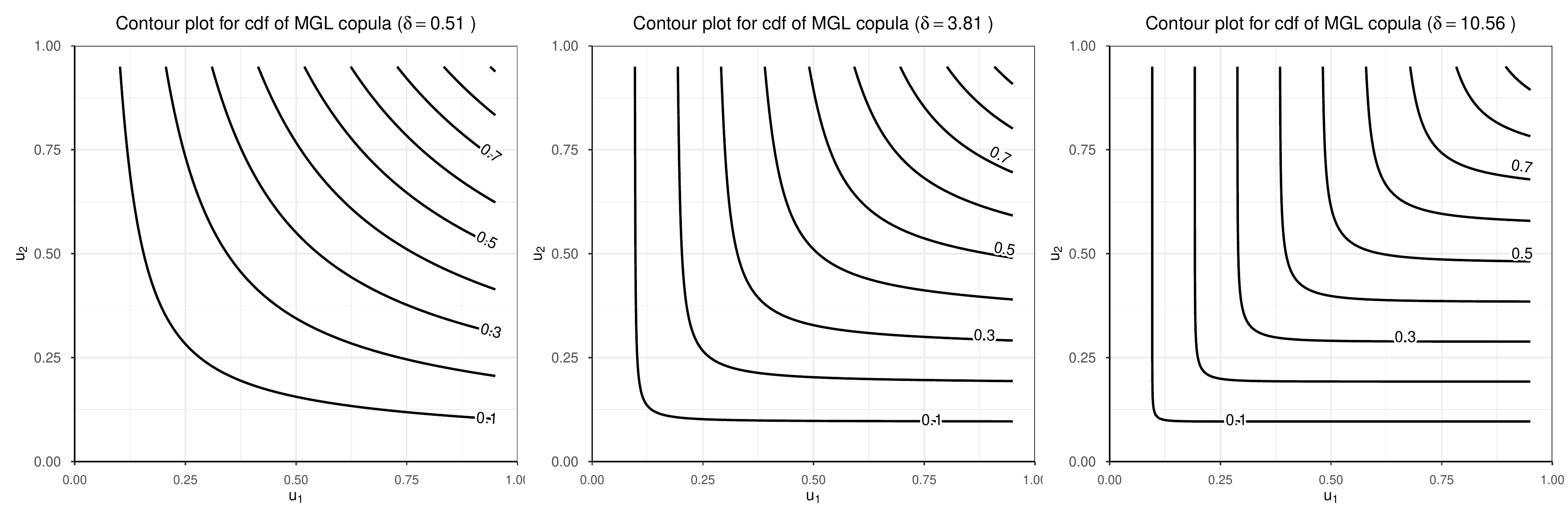}\\
		\includegraphics[scale=0.35]{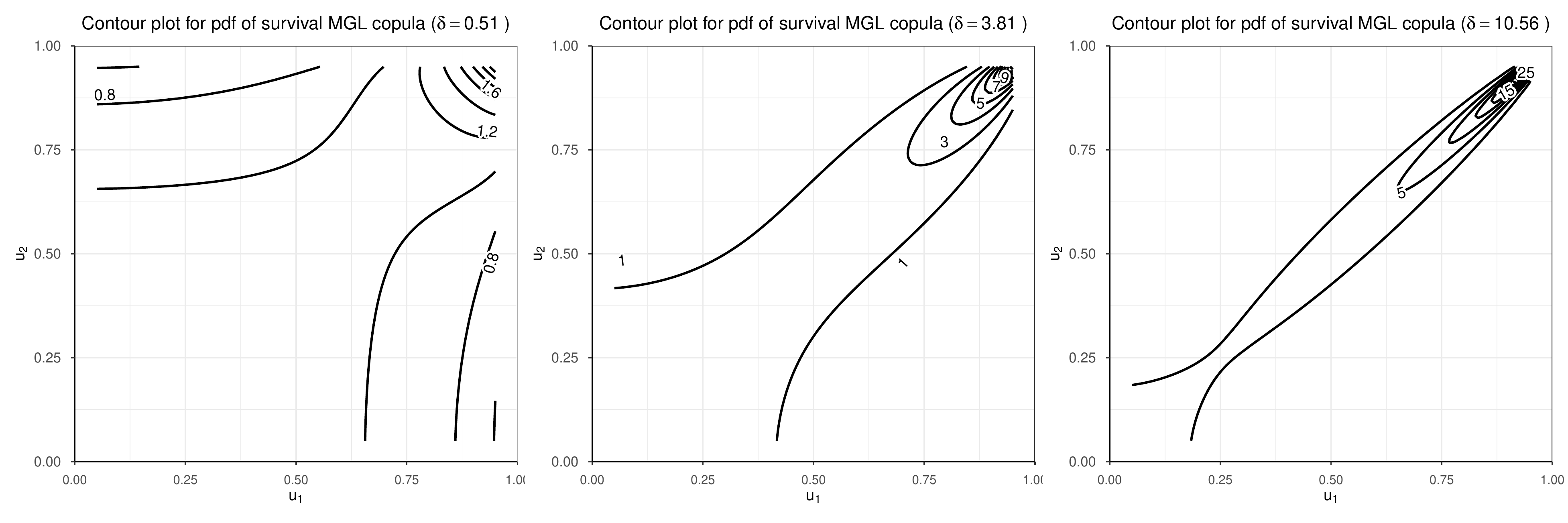}\\
		\includegraphics[scale=0.35]{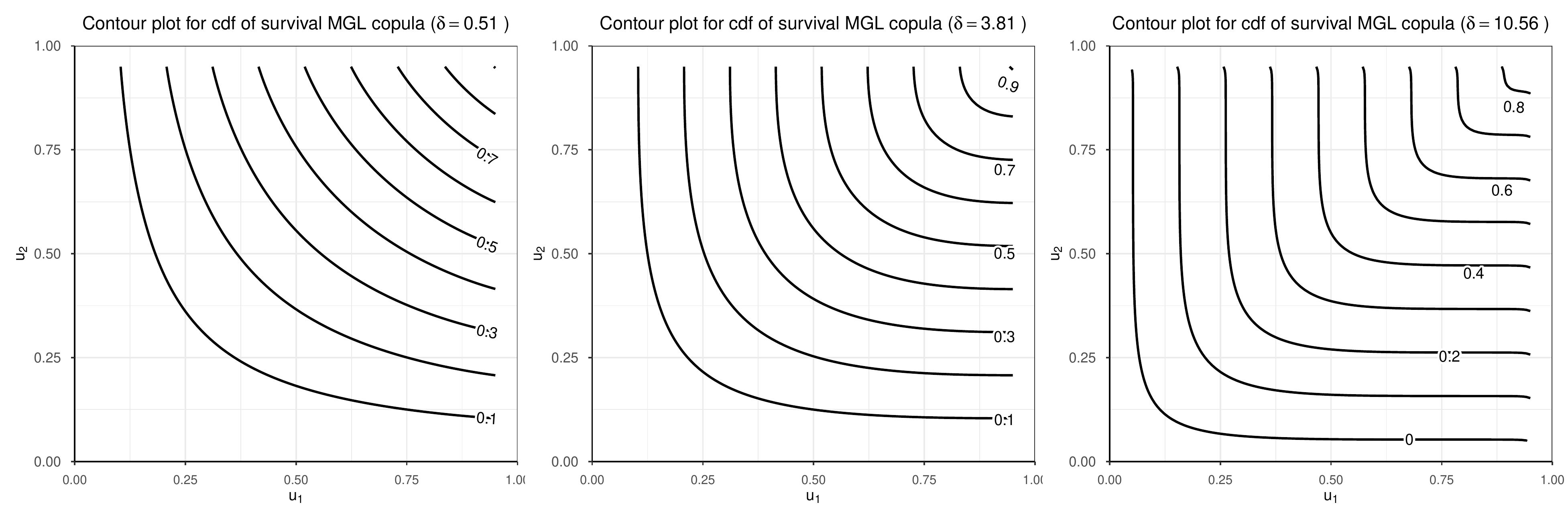}\\
		\caption{
			Contour plots of the joint distribution  and density function of the MGL and survival MGL copula with low, medium and high dependence.
			First column:  low dependence with Kendall's \textit{tau} $= 0.1$,
			second column: medium dependence with Kendall's \textit{tau} $= 0.5$,
			and third column:  high dependence with Kendall's \textit{tau} $= 0.75$.
		}
		\label{fig:counterplot}
	\end{figure}

The next proposition shows that the MGL copula is able to accommodate
		joint extreme events on the lower tail, but not
		on the upper tail.
		The lower tail
		dependence index  reduces to 0 as $\delta$ approaches 0.	
	\begin{proposition}\label{prop:tail-dependence}
		The copula $C^{MGL}$ is asymptotic lower tail  dependent with the indices of  lower and upper tail dependence given respectively by
		\begin{align}
		&\lambda_{l}^{MGL}=2-2I_{\frac{1}{2},\frac{1}{\delta}+\frac{1}{2}}\left(\frac{1}{2}\right), \label{lower tail dependence}\\
		&\lambda_{u}^{MGL}=0. \label{upper tail dependence}
		\end{align}
			\end{proposition}
	Similarly the survival copula $\bar{C}^{MGL}$ allows asymptotic upper tail dependence. 	
In Figure \ref{fig:counterplot} with provide  contour plots with low, medium and high dependence as measured by Kendall's \textit{tau}
for the MGL and survival MGL copula.
The value of parameter $\delta$ describes the
strength of the relationship with
 higher values of $\delta$ implying stronger dependence.
The lower tail dependence and non upper tail dependence are observed in MGL copula, and upper tail dependence is observed with the survival MGL copula. \\

	\noindent
{\bf Extreme-value copula} 
	Here we study the corresponding domain of attraction of the survival MGL copula which is able to capture the upper tail dependence. Following the definition of convergence of dependence structures,  as for instance described in Chapter 8.3.2 in 
	\citet{beirlant2004},
	a copula $C$ is attracted to an extreme value copula $C_0$ 
	if  the limit
		\begin{equation*}
		\lim_{s\to 0}\frac{1-C(1-su_1,..., 1-su_d)}{s}=
		\ell (u_1, ...u_d)
		\end{equation*}
		exists  for all $(u_1, ...u_d)\in
		[0,\infty)^d$. Then  $\ell $ is named the stable tail dependence function. The corresponding copula $C_0$ is then obtained from $\ell$ through	
		\begin{equation}
		-\log C_0\left(\exp(-u_1),...,\exp(-u_d)\right)= \ell (u_1, ...u_d).
		\end{equation}
		\label{eq:extreme value copula}
In the bivariate case $d=2$ the stable tail dependence function $\ell$ can be represented in terms of the Pickands dependence function $A$:
\begin{equation*}
\ell (u_1,u_2) = (u_1+u_2)A\left(\frac{u_2}{u_1+u_2} \right),
\end{equation*}
which is necessarily convex and satisfies the boundary condition $\max(1-w,w)\le A(w)\le 1$.
The extreme value copula is then represented as 
\begin{equation*}
C_0 (u_1,u_2)= \exp \left[ \log (u_1u_2)A\left( \frac{\log u_2}{\log (u_1u_2)}\right)\right],
\end{equation*}
and the upper tail dependence coefficient is given by $\lambda_u=2-2A(1/2)$.
\\
The density function and the $h$-function of the bivariate extreme-value copula are given by
	\begin{align*}
	{c}_0(u_1,u_2)&=	\frac{{C}_0(u_1,u_2)}{u_1u_2}\left[
	\left(
	\frac{\partial \ell(z_1, z_2)}{\partial z_1}\frac{\partial \ell(z_1, z_2)}{\partial z_2}
	\right) - \frac{{\partial^2 \ell(z_1, z_2)}}{\partial z_1 \partial z_2}
	\right]\bigg|_{z_1 = -\log u_1, z_2=-\log u_2},\\
	{h}_{2|1}(u_2|u_1)&= 	\frac{{C}_0(u_1,u_2)}{u_1}\frac{\partial \ell(z_1, z_2)}{\partial z_1}\bigg|_{z_1 = -\log u_1, z_2=-\log u_2},
	\\
	{h}_{1|2}(u_1|u_2) &= \frac{{C}_0(u_1,u_2)}{u_2}\frac{\partial \ell(z_1, z_2)}{\partial z_2}\bigg|_{z_1 = -\log u_1, z_2=-\log u_2}.
	\end{align*}

	\begin{proposition}\label{prop:MGL-EV}
		The extreme value copula  $\bar{C}^{MGL-EV}$ of the  survival MGL copula is given by
		\begin{equation}
		\bar{C}^{MGL-EV}(u_1,u_2;\delta)=\exp\left[\log\left(u_1u_2\right)A_{\delta}\left(\frac{\log\left(u_2\right)}{\log\left(u_1u_2\right)}\right)\right],
		\label{eq:survival MGL-EV1}
		\end{equation}
		where the Pickands dependence function $A_{\delta}$ is given by
		\begin{align}
		A_{\delta}\left(w\right)=w I_{{\frac{1}{2}, \frac{1}{\delta}+\frac{1}{2}}}\left[\frac{\left(1-w \right)^{-\delta}}{\left(1-w \right)^{-\delta} + w^{-\delta} }\right]  + \left(1-w\right) I_{{\frac{1}{2}, \frac{1}{\delta}+\frac{1}{2}}}\left[\frac{w^{-\delta}}{\left(1-w \right)^{-\delta} + w^{-\delta} }\right].
		\label{eq:Pickands}
		\end{align}
	\end{proposition}

	\begin{remark}\label{prop:MGL-EV2}
		The MGL copula also has a limiting lower tail copula given by
		\begin{equation}
		C^{MGL-EV}(u_1, u_2; {\delta})=\frac{(u_1+u_2)\left(1-A_{\delta}\left(\frac{u_1}{u_1+u_2}\right)\right)}{2\left(1 - A_{\delta}\left(\frac{1}{2}\right)\right)},
		\end{equation}
		where $A_{\delta}$ is the Pickands dependence function of $\bar{C}^{MGL-EV}$. 
	\end{remark}

\begin{figure}[htbp]
		\includegraphics[scale=0.38]{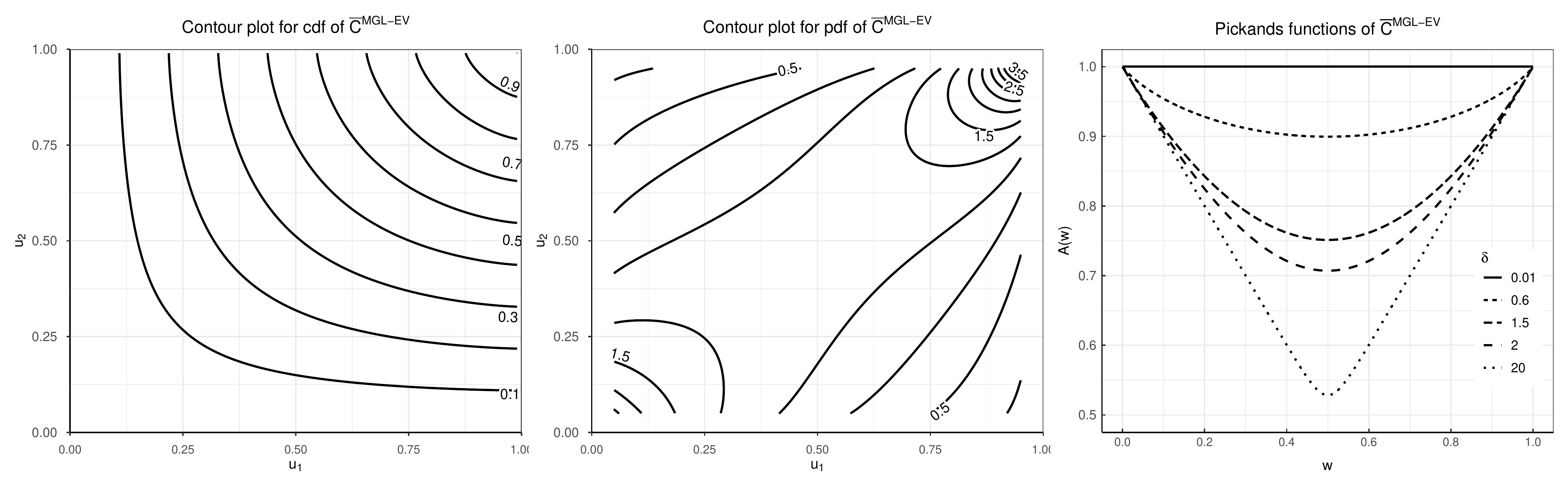} 
		\caption{
			Contour plots with $\delta=1$ for joint distribution and density function of the upper extreme value copula, and the Pickands dependence function  $A_{\delta}$ for different $\delta$ values.
		} 
		\label{fig:counter-evcopula}
	\end{figure}

 Figure \ref{fig:counter-evcopula} displays the contour plots for the joint distribution and density function with $\delta=1$, next to the Pickands dependence function $A_{\delta}$   for different values of the dependence parameter $\delta$. 

%

\section{Copula regressions}\label{section: regression}
The  proposed models can be used to model response variables of any dimension in order to  improve the model fitting by introducing regression analysis that accounts for the dynamic dependence patterns conditioning on different values of covariates.
For regression on a copula parameter,	\cite{hua2014assessing} propose a copula model that incorporates both
regression on each marginal of bivariate response variables and regression on the dependence parameter for the response variables. 
\cite{acar2011dependence}
apply a nonparametric approach for calibrating the dependence parameters according to the covariates, where
the dependence parameter is allowed to change along the covariates.
Here, we assume that the $d$-dimensional pseudo-copula data $(u_1,...,u_d)$ follow 
the survival MGL/survival MGL-EV copula and propose the copula parameter $\delta$ to be modelled as a function of the explanatory variables
\footnote{
For the sake of simplicity, in this section we only show the estimation method for the survival MGL and survival MGL-EV copula  as modelling the upper dependence is often emphasized in actuarial science. The estimation for MGL and MGL-EV copula can be easily obtained in a similar way.
}.
In order to avoid boundary
problems in optimization, we consider a log link function obtaining real values for the copula parameter $\delta_i$:


\begin{align*}
u_{i1},...,u_{id}|\bm{x}_{i}&\sim \text{survival MGL}(\delta_i),\\
\log(\delta_i)& = \bm{x}_{i}^{T}\bm{\beta},
\end{align*}
where $\bm{x}_{i}=(1,x_{i1}...,x_{ik})$ denotes the vector of covariates and $\bm{\beta}=(\beta_0,\beta_{1},...,\beta_k)$ the vector of coefficients.

The pseudo log-likelihood  function  $\log L(;\bm{\beta})$ based on pseudo-copula data $(\bm{u}_{1}, ..., \bm{u}_{d})^T$ for $ i=1,\dots, n$ is given by
\begin{align}
\log L(u_{1}, ..., u_{d}; \bm{\beta}) &= \sum_{i=1}^{n}\log\bar{c}^{\text{MGL}}({u}_{i1}, ..., {u}_{id};\delta_i)\nonumber \\
& = (d-1)\sum_{i=1}^{n}\log \Gamma\left(\frac{1}{\delta_i}\right) + 
\sum_{i=1}^{n}\log\Gamma\left(\frac{1}{\delta_i}+\frac{d}{2}\right) - 
d\sum_{i=1}^{n}\log\Gamma\left(\frac{1}{\delta_i}+\frac{1}{2}\right) \nonumber\\
& \quad \quad + \sum_{i=1}^{n}\left({\frac{1}{\delta_i}+\frac{1}{2}}\right)\sum_{j=1}^{d}\log
\left(
\frac{I^{-1}_{\frac{1}{2},\frac{1}{\delta_i}}({u}_{ij})}{1-I^{-1}_{\frac{1}{2},\frac{1}{\delta_i}}({u}_{ij})}
\right)\nonumber\\
& \quad \quad \quad  - \sum_{i=1}^{n}\left({\frac{1}{\delta_i}+\frac{d}{2}}\right)\log\left(\sum_{j=1}^{d} \frac{I^{-1}_{\frac{1}{2},\frac{1}{\delta_i}}({u}_{ij})}{1-I^{-1}_{\frac{1}{2},\frac{1}{\delta_i}}({u}_{ij})}   + 1\right).
\label{eq: log-likelihood-MGL}
\end{align}

Similarly, the covariates can be also introduced into the parameter $\delta$ in the survival MGL-EV copula:
\begin{align*}
	u_{i1},...,u_{id}|\bm{x}_{i}&\sim \text{survival MGL-EV}(\delta_i),\\
	\log(\delta_i)&=\bm{x}_{i}^{T}\bm{\beta},
\end{align*}
with the pseudo log-likelihood function given by
\begin{align}
\log L(u_{1}, ..., u_{d}; \bm{\beta}) &=\sum_{i=1}^{n}\log \bar{c}^{\text{MGL-EV}}({u}_{i1}, ..., {u}_{id};\delta_i)\nonumber \\
&=\sum_{i=1}^{n}\log \bar{C}^{\text{MGL-EV}}({u}_{i1}, ..., {u}_{id};\delta_i)
	- \sum_{i=1}^{n} \sum_{j=1}^d\log u_j  \nonumber \\
&+  \sum_{i=1}^{n}\log\sum_{m=1}^{d}(-1)^{d-m}\sum_{\pi:\abs{\pi}=m}\prod_{B\in\pi}D_B\ell(z_{i1},...,z_{id})|_{z_1=-\log u_{i1},...,z_d = -\log u_{id}},
\end{align}
where $D_B:=\frac{\partial^{\abs{B}}}{\prod_{j\in B}\partial z_j}$ is defined as the high order partial differentiation operation,
$\pi$ runs through the set of all partitions of the set $\left\{1,...,d\right\}$
and $B\in \pi$ denotes that $B$ runs through the list of all elements of the partition $\pi$,
$\abs{\pi}$ denotes the number of sets in $\pi$. 

For the Newton-Raphson algorithm, the first- and second-order derivatives of 
$\log L(u_{1}, ..., u_{d}; \bm{\beta})$ with respected to $\bm{\beta}$ are required at each iteration.
In our model, there is a closed form for the derivatives for the survival MGL copula, but no simple form for the survival MGL-EV copula is obtained. However they can be obtained numerically, requiring multiple calculations of the log-likelihood.
We show how to calculate the gradient of the log-likelihood \eqref{eq: log-likelihood-MGL}
in the Appendix \ref{App: gradient}. 
The maximum likelihood (ML) estimates are consistent and
asymptotically normal if the copula family is correctly specified. 
The asymptotic variance of the estimates depends on the
estimation method; see sections 5.4, 5.5, and 5.9 in \cite{joe2014dependence} for
a detailed review of asymptotic theory when estimates are
obtained using the joint likelihood, the two-step approach with
parametric margins, and the two-step approach with nonparametric ranks, respectively.

\section{Simulation Study}\label{section: simulation}
	In this section, we check first the accuracy of the ML estimators based on the proposed $d$-dimensional MGL copula regression model discussed in Section \ref{section: regression}
	with respect to sample size $n$.
	We generate $N=1,000$ data sets from $n=100$ to $n= 2,000$ from the $d$-dimensional MGL copula regression model with $k=2$, $\bm{x}_i^T=(1,x_{i1}, x_{i2})$, $\bm{\beta}=(-0.6, 0.5, 0.2)^T$ with the covariates $x_{i1}$ and $x_{i2}$ being generated from the standard normal distribution.
	We consider $d=2$ and $d=10$  to yield  bivariate and high dimensional copula regressions for simulated pseudo-copula data.	

\begin{figure}[htbp]
		\includegraphics[scale=0.45]{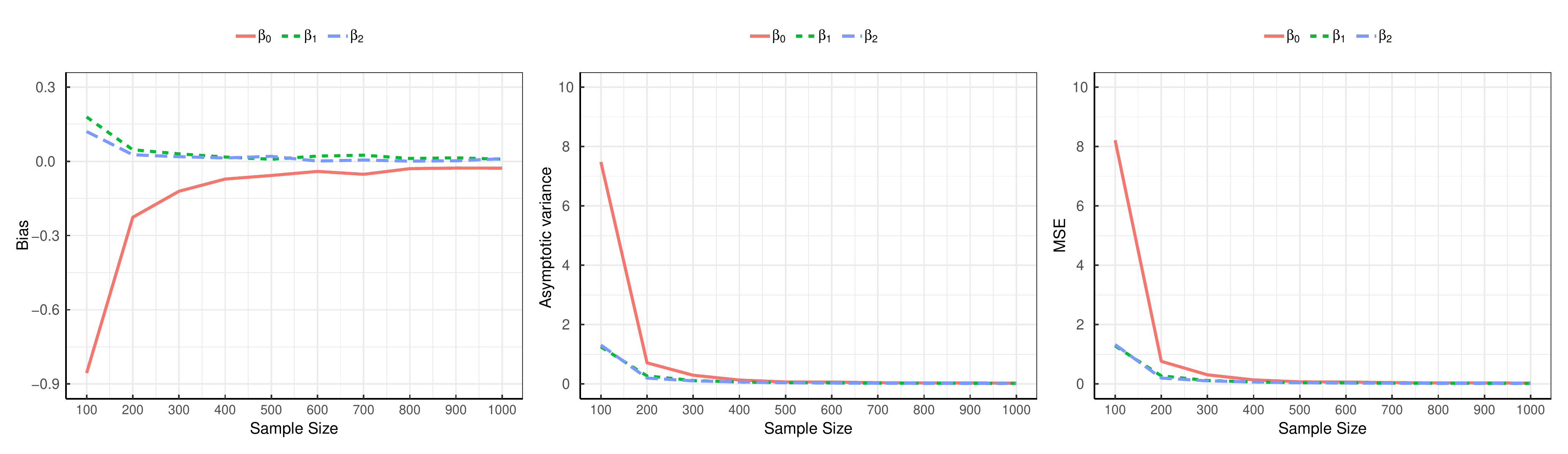}
		\caption{
			Bias (left), asymptotic variance (middle) and MSE (right) of the parameter estimates $(\beta_0,\beta_1,\beta_2)$ for MGL copula regression 
			in case of $d=2$.
			The sample size runs from $n = 100$ to $n = 1,000$. The plots are obtained by
			averaging  over 1,000 samples.}
		\label{fig-simulation-reg-dim2}
	\end{figure}

\begin{figure}[htbp]
			\includegraphics[scale=0.45]{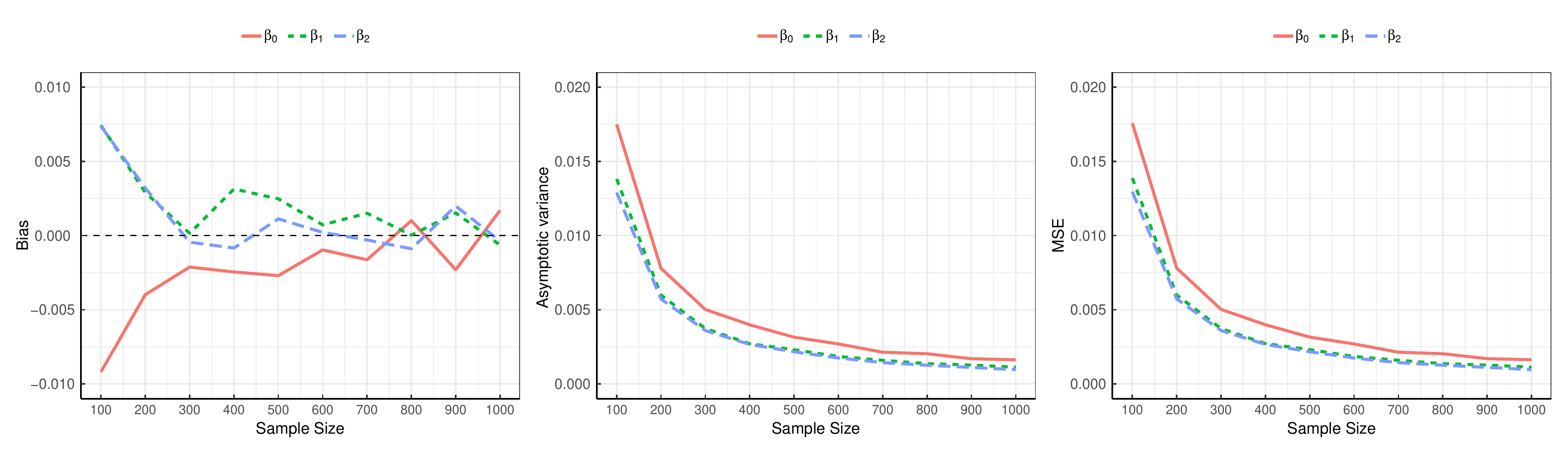}
		\caption{
			Bias (left), asymptotic variance (middle) and MSE (right) of the parameter estimates $(\beta_0,\beta_1,\beta_2)$ for MGL copula regression 
				in case of $d=10$.
				The sample size runs from $n = 100$ to $n = 1,000$. The plots are obtained by
				averaging  over 1,000 samples.
}
		\label{fig-simulation-reg-dim10}
	\end{figure}

	Figures \ref{fig-simulation-reg-dim2} and  \ref{fig-simulation-reg-dim10}
	show how the bias, asymptotic variance and mean squared error (MSE) vary with respect to sample size $n$ in case of $d=2$ and $d=10$ respectively.
	It can be observed that 
estimation of the model parameter $\beta_0$ is less accurate
in smaller samples, while more stability is observed with the estimators of $\beta_1$ and $\beta_2$. 
	As the sample size increases the estimators close up to the true values, with smaller bias, asymptotic variance and MSE.
		
\begin{figure}[htbp]
	\centering
		\includegraphics[scale=0.6]{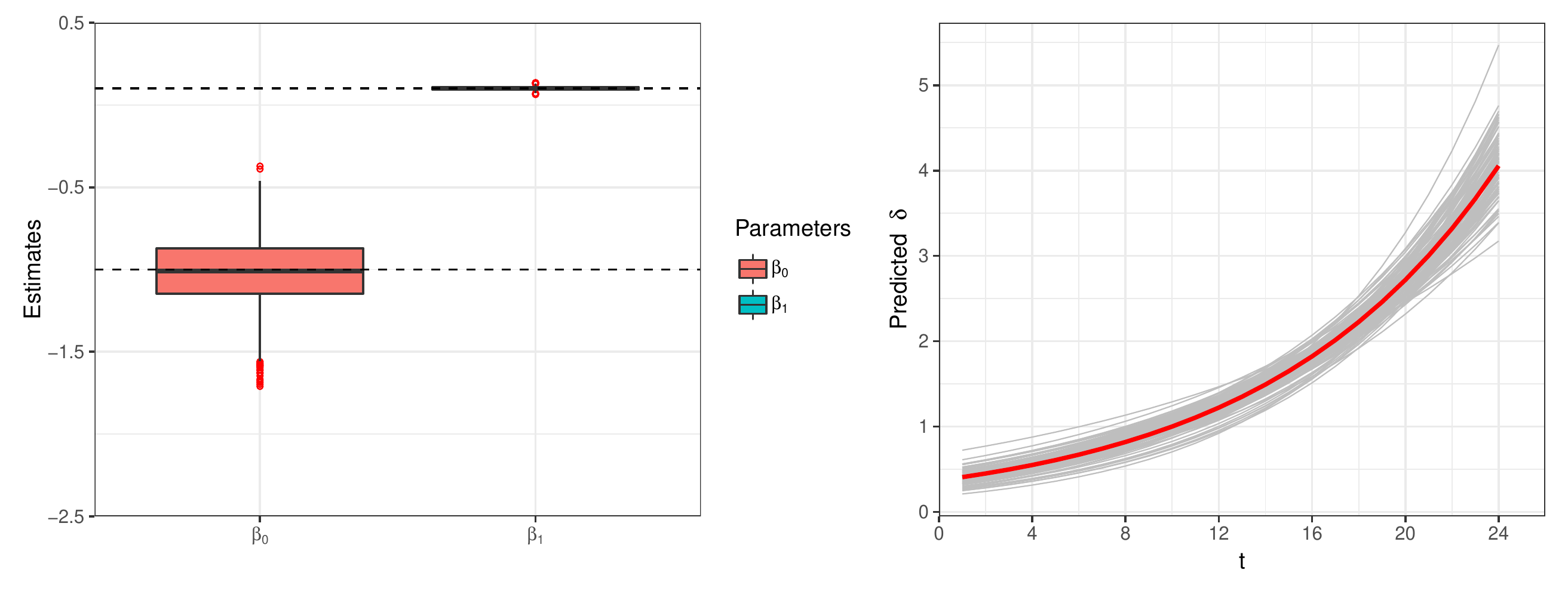}\\
	\caption{
		Boxplots of the parameter estimates from 2,000 survival MGL copula simulated samples of size $n=1,200$ (\textit{left panel}).
		The predictive value of copula parameter $\delta_i$ for different values of the covariate $t$ (\textit{right panel}). 
	}
	\label{fig-simulation-boxplot-predicted}
\end{figure}

\begin{figure}[htbp]
	\centering
	\includegraphics[scale=0.6]{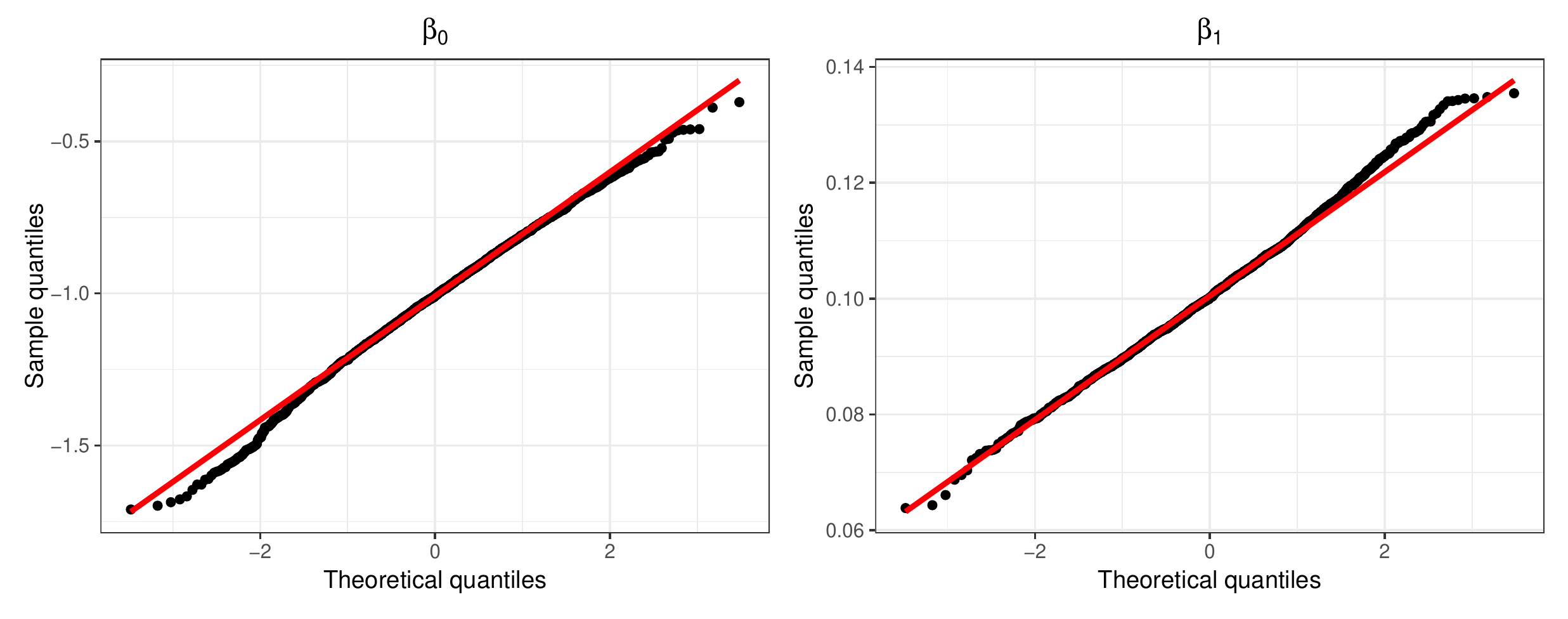}
	\caption{
	Normal QQ plots of the ML parameter estimates from the survival MGL copula regression simulations with sample size $n=1,200$. 
	}
	\label{fig-simulation-qqpplot}
\end{figure}

	Dynamic dependence modelling has been a popular research topic in actuarial science \citep{hua2014assessing}, with special emphasis to upper tail dependence.
	In order to study dynamic upper tail dependence structures, we apply the copula regression model to simulated bivariate data using the survival MGL copula with a time covariate.
	We generated  random samples of size $n=1,200$ from the bivariate survival MGL copula, and the sample size for each time point.
	The dependence parameter is assumed to be a function of the claim duration in months $(t \in \left\{1,...,24\right\})$ according to 
$\delta_{t}=\exp(-1+0.1 t)$. 
Figure \ref{fig-simulation-boxplot-predicted} presents the boxplots of the parameter estimates from 2,000 Monte Carlo simulations.
The median estimates of $\beta_0$ and $\beta_1$  are very close to the true values.
To demonstrate the approximate normality of the estimators 
the normal QQ plots of the estimated parameters are given in Figure \ref{fig-simulation-qqpplot}, which show acceptable results. 
Figure \ref{fig-simulation-boxplot-predicted} shows the predictive curve between the copula parameters $\delta_t$ and duration $t$.
The red line shows the dependence pattern from the true model.
The gray lines are generated based on the 
the estimates of the copula parameter $\hat{\delta}_i$ from 2,000 Monte Carlo simulations.
It can be observed that the dependence in the upper tail is close to independence at smaller $t$ and increases with $t$.

\newpage
	\section{Real-data illustrations}\label{section:application}
	We now illustrate the proposed methods with  two practical examples which are investigated without and with covariates.

	\subsection{Danish fire insurance data}
As the first example, we fit the bivariate copula and  regression models to 
the Danish fire insurance data set which was
collected from the Copenhagen Reinsurance Company and comprises 2167 fire losses over the period 1980-1990.
The claims have been adjusted for inflation to reflect 1985 values and are expressed in millions of Danish Krone and can be found in the R package: \texttt{fitdistrplus}.
The total claims in the multivariate data set is divided into  building loss, contents loss and  profit loss.
This data set was already analyzed in \cite{hashorva2017some} and \cite{lu2021nonparametric} among others.
Here we model the dependence between  building loss and  contents  loss, and we consider the observations where both components are non-zero.
There is a total of $n=1502$ observations that are positive in both variables.

\begin{figure}[htbp]
	\centering
	\includegraphics[scale=0.6]{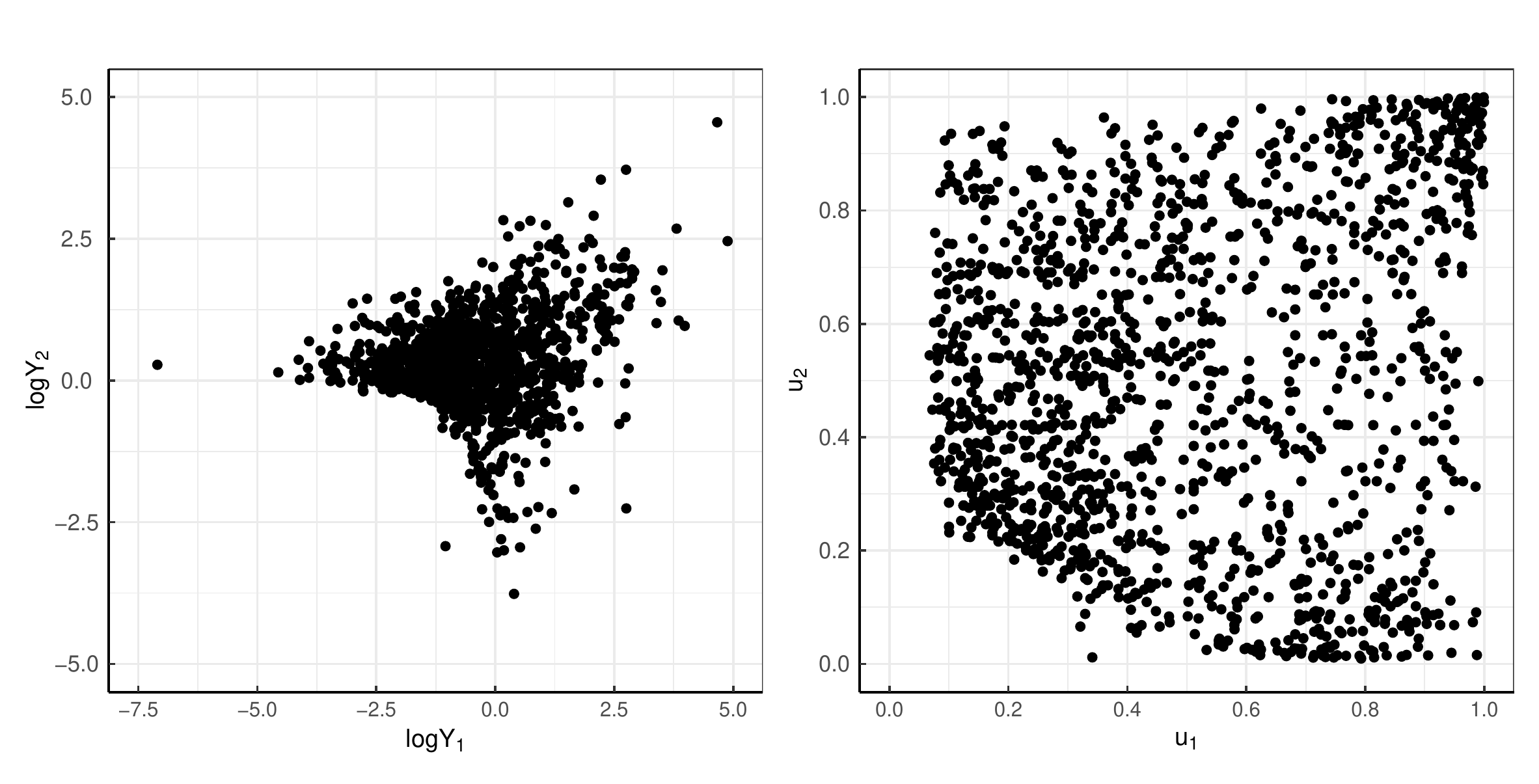} 
	\caption{Scatter plots of the log transformed data  (\textit{left panel}) and the pseudo copula data based on  the nonparametric kernel smoothing method (\textit{right panel}).}
	\label{fig: danish-scatter-plot}
\end{figure}

Figure \ref{fig: danish-scatter-plot} displays the scatter plot of the log transformed data $(Y_{i1}, Y_{i2})$ and of the pseudo-copula data $(u_{i1}, u_{i2})$ ($i=1,...n$) based on the kernel smoothing method \begin{align*}
	u_{ij}&=\hat{F}_n(y_{ij})=\int_{-\infty}^{y_{ij}}\frac{1}{n}\sum_{j=1}^{n}K_h(Y_{ij}-u)du,\quad \text{for} \quad j=1,2,
\end{align*}
where $K(\cdot)$ is a kernel function and $K_h=K(\cdot/h)/h$. Here we have chosen the standard  Gaussian kernel and $h=0.2$.   From these positive right upper tail dependence appears.
The empirical value of Kendall's \textit{tau} equals 0.085.

Table \ref{tab:estimates - danish} reports the estimation results, AIC and BIC values of the survival MGL and the survival MGL-EV copula, along with four other families of copulas with
positive upper tail indices,  the MGB2 copula discussed in \cite{yang2011generalized},
the Gumbel copula,  the Student $t$ copula, and the Gaussian copula.
The Gumbel copula is an extreme-value copula and also belongs to the
Archimedean family, whereas the Student $t$ copula and Gaussian copula belongs to the elliptical copulas.
We estimate the copula parameters $\bm{\gamma}$.
\begin{itemize}
	\item For the survival MGL, $\bm{\gamma}=\delta$,
	\item  For the survival MGL-EV, $\bm{\gamma}=\delta$,
	\item
	for the MGB2
	copula, $\bm{\gamma}$ is a 3-vector of $(p_1, p_2, q)^T$,
	\item for the Gumbel copula,
	$\bm{\gamma}=\delta$,
	\item for the Gaussian copula, $\bm{\gamma}=\rho$,
	\item
	for the Student $t$ copula, $\bm{\gamma}= (\rho, v)^T$ and the degree of freedom $v$ is determined by ML.
\end{itemize}
In terms of the AIC and BIC values, the MGB2 and survival MGL are preferred over the other four families of copulas. 
In addition, in order to analyze the model fitting in upper and lower regions, we
consider the squared fit error over a  region $A\subseteq \left[0,1\right]$, defined as
(see \cite{li2014distorted})
\[
e_A(C) =\left(
\frac{1}{m(A)}\int \int_{A} \abs{C(u,v)-C^{emp}(u,v)}^2 du dv
\right)^{\frac{1}{2}},
\]
where $m(A)$ is the Lebesgue measure of a set $A$, $C$  the fitted copula and $C^{emp}$ is the empirical copula which is defined as  $C^{emp}(t_1,t_2)=\sum_{i=1}^{n}\bm{1}_{y_{i1}<t_1}\bm{1}_{y_{i2}<t_2}/n$.
It can be observed that the survival MGL copula possesses the better performance over the regions  $\left[0,0.05\right]^2$ and  $\left[0.95,1\right]^2$.

To further investigate the tail behavior of the proposed models, we also consider the tail-weighted measures of dependence proposed by \cite{krupskii2015tail}
and applied in \cite{krupskii2018factor}. 
The measures provide useful tools for summarizing the strength of dependence in different joint tails for each pair of variables with value close to 0 or 1 corresponding to very weak or strong dependence in the tails respectively.
Unlike the goodness-of-fit procedures such as AIC and BIC statistics, the tail-weighted measures of dependence can be used as additional scalar measures to distinguish bivariate copulas with roughly the same overall monotone dependence, as well as for assessing the adequacy of fit of multivariate copulas in the tails.\\
The empirical and model-based tail-weighted measures of dependence in the upper tail are given respectively by: 
\begin{align*}
\varrho_{U}(a,p)&=\widehat{\text{Cor}}\left[
a\left(1 - \frac{1-R_{i1}}{p}\right), a\left(1 - \frac{1-R_{i2}}{p}\right)
\bigg| 1 - R_{i1}<p,  1 - R_{i2}<p
\right],\\
\rho_{U}(a,p;C)&=\frac{C(p,p)m_{12} - m_1m_2}{
\left\{
\left[C(p,p)m_{11}-m_1^2\right]
\left[C(p,p)m_{22}-m_2^2\right]
\right\}^{1/2}
},
\end{align*}
where the notation $\widehat{\text{Cor}}\left[y_{i1}, y_{i2}|(y_{i1}, y_{i2})\in B\right]$ is  shorthand for
\[\frac{\sum_{i\in J_{B}}y_{i1}y_{i2} -   n_{B}^{-1}\sum_{i\in J_{B}}y_{i1}\sum_{i\in J_{B}}y_{i2}  }{
\left[\sum_{i\in J_{B}}y_{i1}^2 - n_B^{-1}\left(\sum_{i\in J_{B}}y_{i1}\right)^2\right]^{1/2}
\left[\sum_{i\in J_{B}}y_{i2}^2 - n_B^{-1}\left(\sum_{i\in J_{B}}y_{i2}\right)^2\right]^{1/2}
},
\]
with $J_B=\left\{i: (y_{i1}, y_{i2}\in B)\right\}$, $n_B$ is the cardinality of $J_B$, and
\begin{align*}
	m_{12}&=\frac{1}{p^2}\int_{0}^p\int_{0}^p a'\left(1-\frac{u_1}{p}\right)a'\left(1-\frac{u_2}{p}\right)C(u_1, u_2)du_1du_2,\\
	m_1 &= \frac{1}{p}\int_{0}^pa'\left(1-\frac{u_1}{p}\right)C(u_1,p)du_1,\quad m_2 = \frac{1}{p}\int_{0}^pa'\left(1-\frac{u_2}{p}\right)C(p,u_2)du_2,\\
	m_{11} &= \frac{1}{p}\int_{0}^p2a\left(1-\frac{u_1}{p}\right)a'\left(1-\frac{u_1}{p}\right)C(u_1,p)du_1,\\
	m_{22}&= \frac{1}{p}\int_{0}^p2a\left(1-\frac{u_2}{p}\right)a'\left(1-\frac{u_2}{p}\right)C(p,u_2)du_2,
\end{align*}
where $a(\cdot)$ is a weighting function, 
$a'(\cdot)$ is the first-order derivative,
$p$ a truncation level, and $C$ a fitted copula.
In this case, we follow the method in \cite{krupskii2015tail} by using the power function for $a(u)=u^k$ with $k=6$ and the truncation level $p=0.5$. 
The power function with large $k$ puts more weight in the tail.
The results of the power function for $k=5$ and $k=7$ are also investigated for comparison.
Table \ref{tab:weight - tailed - danish} reports the estimation results of 
tail-weighted measures of  dependence for the six families of copulas.
To account for variability in parameter estimates, we use the bootstrap to construct 95\% confidence intervals. 
One can see that the model-based estimates of tail-weighted measures of dependence 
are quite close to the empirical estimates for survival MGL copula.
The MGB2 copula model slightly overestimates the empirical upper tail dependence,
while the Gumbel model slightly underestimates the empirical $\hat{\varrho}_{U}(a;p=0.5)$ for different values of $k$.
Moreover, the narrow confidence intervals for the tail-weighted measures of dependence  indicates that the survival MGL model is appropriate for modelling dependence in the upper tails for this case.

\begin{table}[htbp]
	\centering
	\caption{Estimates and goodness fit of candidate of copulas for Danish fire insurance data set.}
			\setlength{\tabcolsep}{1.5mm}{
	\begin{threeparttable}
			\begin{tabular}{lcccccccc}
			\toprule
			\multirow{2}[0]{*}{Copula} & \multicolumn{3}{c}{\multirow{2}[0]{*}{Estimates and standard errors}} & \multirow{2}[0]{*}{Loglike} & \multirow{2}[0]{*}{AIC} & \multirow{2}[0]{*}{BIC} & \multirow{2}[0]{*}{$e_{\left[0,0.05\right]^2}$} &\multirow{2}[0]{*}{$e_{\left[0.95,1\right]^2}$}\\
			& \multicolumn{3}{c}{}  &       &       &  &&\\
			\hline
			\multirow{2}[0]{*}{Gaussian} & $\hat{\rho}=0.252$ &    -   &    -   & \multirow{2}[0]{*}{35.60} & \multirow{2}[0]{*}{-69.20} & \multirow{2}[0]{*}{-63.89}& \multirow{2}[0]{*}{2.23}&
			\multirow{2}[0]{*}{103.03}
			\\
			& (0.027) &  -     &   -    &       &       &  \\
			\multirow{2}[0]{*}{Student $t$} & $\hat{\rho}=0.193 $ & $\hat{v}=3.400$ &  -     & \multirow{2}[0]{*}{64.08} & \multirow{2}[0]{*}{-124.17} & \multirow{2}[0]{*}{-113.54}  &\multirow{2}[0]{*}{9.34}& \multirow{2}[0]{*}{88.39}\\
			& (0.032) & (0.457) &    -   &       &       &  \\
			\multirow{2}[0]{*}{Gumbel} &$\hat{\delta}= 1.211$ &    -   &   -    & \multirow{2}[0]{*}{79.13} & \multirow{2}[0]{*}{ -156.25} & \multirow{2}[0]{*}{-150.94} &  \multirow{2}[0]{*}{0.79}& \multirow{2}[0]{*}{1.56}\\
			& (0.022) &   -    &  -     &       &       &  & & \\
			{Survival MGL} & $\hat{\delta}=0.892$ &  -     &  -     & \multirow{2}[0]{*}{{115.97}} & \multirow{2}[0]{*}{{ -229.93}} & \multirow{2}[0]{*}{{-224.62}} &  \multirow{2}[0]{*}{\textbf{0.27}} & \multirow{2}[0]{*}{{\textbf{0.54}}}
			\\
			 & (0.067) &  -     &  -     &       &       &  \\
			Survival MGL-EV & $\hat{\delta}=0.655$ & - & - & \multirow{2}[0]{*}{81.76} & \multirow{2}[0]{*}{-161.52 } & \multirow{2}[0]{*}{{-156.20}}& \multirow{2}[0]{*}{0.79} & \multirow{2}[0]{*}{{1.55}}\\
				& (0.040) & - & - &       &       &  \\
			\multirow{2}[0]{*}{MGB2} & $\hat{p_1}=0.233$ & $\hat{p_2}=1.123$ & $\hat{q}=0.939$ & \multirow{2}[0]{*}{\textbf{127.82}} & \multirow{2}[0]{*}{\textbf{-249.64}} & \multirow{2}[0]{*}{\textbf{-233.69}} &   \multirow{2}[0]{*}{0.28} & \multirow{2}[0]{*}{{0.80}}\\
			& (0.046) & (0.561) & (0.209) &       &       &  \\
			\bottomrule
		\end{tabular}%
		\begin{tablenotes}
			\footnotesize
			\item Notes: The square fit error $e_{A}(C)$ in regions $A = \left[0,0.05\right]^2$ and $A = \left[0.95,1\right]^2$ are rescaled by $\times 10^8$.
			The standard error is reported in······ brackets.
		\end{tablenotes}
	\end{threeparttable}
}
	\label{tab:estimates - danish}%
\end{table}%

\begin{table}[htbp]
	\centering
	\caption{Estimates of empirical and fitted tail-weighted measures of dependence in the upper tail ($\varrho_U(a;p)$ and $\rho_{U}(a;p)$) and the model-based 95\% confidence intervals for the Danish fire insurance data set.}
		\setlength{\tabcolsep}{6mm}{
	\begin{threeparttable}
			\begin{tabular}{lccc}
			\toprule
			\multirow{2}[0]{*}{Models} & Estimates & Estimates & Estimates \\
			& $a(u)=u^5,p=0.5$   & $a(u)=u^6,p=0.5$   & $a(u)=u^7,p=0.5$ \\
			\hline
			\multirow{2}[0]{*}{Empirical} & 0.434 & 0.427 & 0.419 \\
			& (0.357,0.503) & (0.348,0.499) & (0.339,0.492) \\
			\multirow{2}[0]{*}{Gaussian} & 0.101 & 0.099 & 0.097 \\
			& (0.078,0.123) & (0.076,0.121) & (0.074,0.118) \\
			\multirow{2}[0]{*}{Student t} & 0.352 & 0.360  & 0.366 \\
			& (0.276,0.364) & (0.282,0.372) & (0.286,0.378) \\
			\multirow{2}[0]{*}{Gumbel} & 0.317 & 0.324 & 0.330 \\
		&	(0.290,0.344) & (0.297,0.352) & (0.303,0.358) \\
			\multirow{2}[0]{*}{Survival MGL} & 0.420  & 0.429 & 0.436 \\
		&\textbf{	(0.392,0.448) }& \textbf{(0.400,0.458)} &\textbf{ (0.406,0.465) }\\
			\multirow{2}[0]{*}{Survival MGL-EV} & 0.306 & 0.314 & 0.319 \\
		& (0.280,0.333) & (0.287,0.340) & (0.292,0.346) \\
			\multirow{2}[0]{*}{MGB2} & 0.434 & 0.444 & 0.452 \\
		& \textbf{(0.378,0.487)} & \textbf{(0.387,0.498)} & \textbf{(0.393,0.506) }\\
			\bottomrule
		\end{tabular}  
		\begin{tablenotes}
			\footnotesize
			\item Notes: The 95\% confidence intervals are reported in brackets and the confidence intervals that contain the empirical value are shown in bold font.
			The results are baesd on 200 bootstrap samples
		\end{tablenotes}
	\end{threeparttable}
}
	\label{tab:weight - tailed - danish}%
\end{table}%

We further investigate the dynamic dependence introducing the covariate \textit{{Year}} into the dependence parameter in the survival MGL and survival MGL-EV regression model respectively.
The natural cubic splines are used to allow flexible relationships between \textit{{Year}} and the dependence parameter.
The log link function is considered for the dependence parameter $\delta_i$:
\begin{equation*}
	\log \delta_{i}=\text{ns}_i(\text{Year})=\beta_1b_1(\text{Year})+...+\beta_kb_k(\text{Year}),
\end{equation*}
where $\text{ns}_i(x)$ denote the natural cubic splines, with $b_1(x),...,b_p(x)$ denoting the spline basis and $\beta_1,...,\beta_k$  the regression coefficients. 
We have chosen the 50\% percentile of \textit{Year} as one knot for the natural cubic spline and there are three coefficients $\beta_{k}, k = 1,...,3$ to be estimated in the copula regression.  

\begin{figure}[htbp]
	\centering
	\includegraphics[scale=0.55]{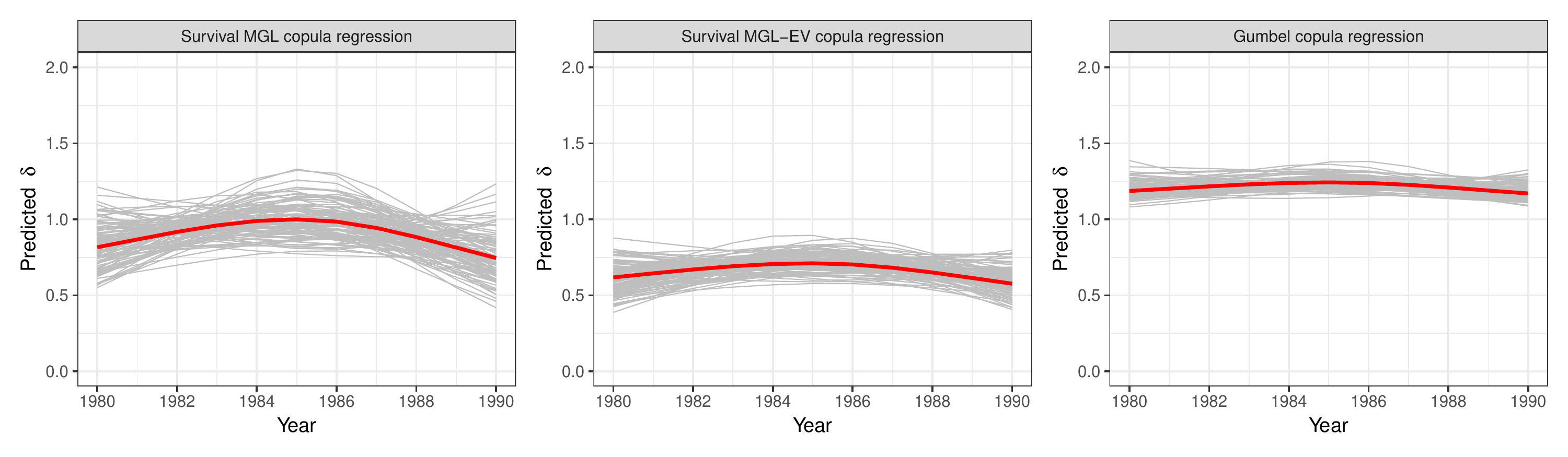} 
	\caption{The predicted value of copula parameter with different value of the covariate \textit{Year} for the Danish fire insurance data set.
	The gray lines are generated based on the simulated coefficients for the natural cubic spline, and their values are generated using a multivariate normal distribution with mean equal to the  MLE's  $\hat\beta_k$ for $k=1,2,3$, and the covariance matrix being the inverse of the Hessian matrix.}
	\label{fig-danish-predicted}
\end{figure}

{Table} \ref{tab:estimates - reg - danish} reports the estimates and standard errors of the regression coefficients, together with the log-likelihood and information statistics of the survival MGL and 
survival MGL-EV copula regression model. The estimation results are also reported for the Gumbel regression model. 
In terms of the AIC and BIC values 
it can be observed that the survival MGL copula provide a better overall fitting than the Gumbel regression.
Figure \ref{fig-danish-predicted} presents the relationship between \textit{Year} and the dependence parameter.
A non-linear relationship appears with an initial increasing  dependence followed by a final decrease.

	\begin{table}[htbp]
	\centering
	\caption{Estimates and goodness fit for the survival MGL and Gumbel regression models for the Danish fire insurance data set.}
	\setlength{\tabcolsep}{5mm}{
		\begin{threeparttable}
			\begin{tabular}{c|cccccc}
				\toprule
				\multicolumn{1}{c|}{\multirow{2}[0]{*}{Parameters	}} 	 & \multicolumn{2}{c}{Survival MGL} & \multicolumn{2}{c}{Survival MGL-EV}  &\multicolumn{2}{c}{Gumbel} \\
				& Estimates & S.E.   & Estimates & S.E. & Estimates & S.E.\\
				\hline	
		$\beta_1$ & 		0.205 &      0.243 &     -0.022&   0.194 & -0.422 &      0.341 \\			
			$\beta_2$ & 	-0.319 &      0.186 &     -0.968 & 0.148& -3.483 &      0.266 \\			
			$\beta_3$ & 	-0.191 &      0.225 &    -0.192& 0.174& -0.446 &      0.316 \\	
				\hline
				Loglike & \multicolumn{2}{c}{\textbf{116.69}} & \multicolumn{2}{c}{{82.34}} & \multicolumn{2}{c}{79.69} \\
				AIC   & \multicolumn{2}{c}{\textbf{-227.37}} & \multicolumn{2}{c}{{-158.69}} & \multicolumn{2}{c}{-153.39} \\
				BIC   & \multicolumn{2}{c}{\textbf{-211.43}} & \multicolumn{2}{c}{{-142.75}} & \multicolumn{2}{c}{-137.44} \\
				\bottomrule
			\end{tabular}%
			\begin{tablenotes}
				\footnotesize
				\item Notes: In order to avoid boundary problem in MLE procedures, 
				we consider a log link function obtaining real values for Gumbel copula regression:
				$\log(\delta_{i}-1)=\text{ns}_i(\text{Year})$ for all $\delta_i>1$.
			\end{tablenotes}
		\end{threeparttable}
	}
	\label{tab:estimates - reg - danish}%
\end{table}%

	\subsection{Chinese earthquake loss data}
	As the second example,
	we consider an earthquake loss data set concerning the Chinese mainland, which contains risk information on 291 earthquake events with magnitude greater than 4.0 from 1990 to 2015.
	The data set is collected from the ``China earthquake yearbook" and also analyzed in \cite{li2019jan}.
	The data set contains the occurrence time, location, the number of casualties and the total economic loss of each earthquake event\footnote{One earthquake event is defined as a earthquake resulting in one of the damage types, such as casualties, economic losses, and damage to buildings.}. Among them, casualties are defined as fatalities and injured people, which are due to damage to occupied buildings.
	Table \ref{tab:earthquakes datset} reports the  major earthquake disasters in China since 1990.
	The total economic damage is expressed in millions of Chinese Yuan (CNY) and are adjusted for inflation to reflect values in 2015.
	In particular, the 2008 earthquake in Sichuan is the most damaging earthquake. {It did cost about 69,227 lives and caused 845.11 billion CNY direct total economic damage.}

	Here we focus on the dependence structure between the total economic losses and the number of casualties. The pairwise Kendall's \textit{tau} is 0.548 and the Spearman's \textit{rho} 0.704.
	The total economic loss variable is of continuous nature and is defined as the positive economic loss associated with an earthquake impact as determined in the
	weeks and sometimes months after the event.
	The number of casualties are semi-continuous data with heavy tails which are used to measure the earthquake risks associated with the fatalities and injured people.
	The kurtosis and skewness of these two variables indicate a heavy tailed nature of these variables.
			 36.08\% of the earthquake events showed no casualties and so the casualties variable exhibits
	 over-dispersion  with a significant fraction of zero observations.

	\begin{table}[htbp]
		\centering
		\caption{Major Chinese earthquakes since 1990.}
		\begin{tabular*}{\hsize}{@{}@{\extracolsep{\fill}}ccccccc@{}}
			\toprule
			\tabincell{c}{Time of\\Occurrence} & Location & Magnitude & \tabincell{c}{Deaths} & \tabincell{c}{Injuries} &\tabincell{c}{The number \\of casualties} &\tabincell{c}{	The total \\economic losses} \\
			\hline
			2013/08/31&Yunnan &5.9&3&63&66&2834\\
			2012/06/30 & Xinjiang & 6.6       & 0     & 52    & 52&1967
			\\
			2013/11/23 & Jilin & 5.5        & 0     & 25    & 25& 1990
			\\
			2005/11/26 & Jiangxi & 5.7       & 13    & 775   & 788 &2023
			\\
			2009/07/09 & Yunnan & 6.0       & 1     & 372   & 373 &2154
			\\
			2014/12/06 & Yunnan & 5.9       & 1     & 22    & 23& 2377
			\\
			2011/03/10 & Yunnan & 5.8     & 25    & 314   & 339 &2385
			\\
			1996/02/03 & Yunnan & 7.0         & 309   & 17057 & 17366 &2500
			\\
			2013/08/12 & Tibet & 6.1       & 0     & 87    & 87&2707
			\\
			2014/11/22 & Sichuan & 6.3      & 5     & 78    & 83& 4232
			\\
			2008/08/30 & Sichuan & 6.1       & 41    & 1010  & 1051& 4462
			\\
			2012/09/07 & Yunnan & 5.7        & 81    & 834   & 915&4771
			\\
			2014/10/07 & Yunnan & 6.6       & 1     & 331   & 332 &5110
			\\
			2015/07/03 & Xinjiang & 6.5      & 3     & 260   & 263 &5430
			\\
			2015/04/25 & Tibet & 8.1     & 27    & 860   & 887&10302
			\\
			2014/08/03 & Yunnan & 6.5        & 617   & 3143  & 3760& 19849
			\\
			2010/04/14 & Qinghai & 7.1       & 2698  & 11000 & 13698& 22847
			\\
			2013/07/22 & Gansu & 6.6       & 95    & 2414  & 2509 &24416
			\\
			2013/04/20 & Sichuan & 7.0         & 196   & 13019 & 13215& 66514
			\\
			2008/05/12 & Sichuan & 8.0        & 69227 & 375783 & 445010& 845110
			\\
			\bottomrule
		\end{tabular*}%
		\label{tab:earthquakes datset}%
	\end{table}%

	For the continuous total economic loss outcome, we consider the univariate heavy-tailed GLMGA distribution as proposed in \cite{li2019jan}.
Three other competitive heavy-tailed distributions, namely,
	log-gamma, Fr\' echet,
	GlogM \citep{bhati2018generalized}, and double-Pareto-Lognormal (DPLN) distribution \citep{reed2004double} are discussed  in details for comparison. 	
	The estimation results and model selections are reported in Appendix \ref{App: GLMGA}.

	For the semi-continuous number of casualties variable $Y_2$ we fit a composite model. 	The composite model assumes  a threshold $u$ below which $Y_2$ is  modelled using a count distribution, whereas  above $u$  a heavy-tailed distribution such as the generalized Pareto (GP) distribution with location parameter $\mu\in \mathbb{R}$, shape parameter $\sigma >0$ and scale parameter $\xi\in \mathbb{R}$ \citep{pickands1975statistical} can be used.
	It is a usual procedure in actuarial science to combine two distributions in a so-called splicing or composite
	model, see e.g. \cite{bakar2015modeling}, \cite{Leppisaari2016Modeling}, and \cite{grun2019extending}.

	The density  and distribution function of a two-component model for $Y_2$ is expressed as
	\begin{align}\label{eq:f2}
	f_{Y_2}(y)&= \begin{cases}
	\omega f_{Y_2}^{d}(y), & y\le u,\\
	(1-\omega) f_{Y_2}^{c}(y), & y> u,
	\end{cases}
	\end{align}
	\begin{align}\label{eq:F2}
	F_{Y_2}(y) &= \begin{cases}
	\omega F_{Y_2}^{d}(y), & y\le u,\\
	\omega + (1-\omega)F_{Y_2}^{c}(y), & y> u,\\
	\end{cases}
	\end{align}
	where $f_{Y_2}^d$ and $F_{Y_2}^d$ denote the density and  cdf of the corresponding right-truncated count distribution, and $f_{Y_2}^c$ and $F_{Y_2}^c$ of the heavy-tailed continuous component.
	In the truncated count component, 
	we consider the right-truncated negative binomial Type II distribution (NBII) with  mean parameter $ \lambda$ and dispersion parameter $\phi$ with the variance being $1+\phi\lambda^2$.
		For more details on truncated negative binomial regression and likelihood functions, see e.g. \cite{shi2015dependent}.
	 	In practical applications one usually tries to set the
	threshold as low as possible, subject to the GP distribution providing an acceptable fit.
	Here a threshold value of $u=20$	is chosen.
	For estimation purpose, we  fix the mixing weight $w$  which can be expressed as $\hat{w}=(n-n_c)/n$ by assuming the exceedance times of the threshold $u$ to follow a homogeneous Poisson process,
	where $n$ is the total number of earthquake events and $n_c$ is the number of observed exceedances over the threshold $u$  \citep{Leppisaari2016Modeling}.
	There are a total of 194 observations in the count data set exceeding the threshold $u =
	20$. Therefore, the estimated for weight $\hat{w}=194/291=66.67\%$ for the casualties data.
	Moreover, the parameters in right-truncated negative binomial distribution
	are estimated by using MLE with $\hat{\lambda}=37.42$ and $\hat{\phi} =  5.45$.
	The ML estimates for GP distribution is 
	with the location parameter $\mu=20$, the shape $\hat{\sigma}=1.87$, and the scale $\hat{\xi}=56.58$. 
	The results for the goodness fit of the count distribution can be found in Appendix \ref{App: casualties}.\\

\begin{figure}[htbp]
		\centering
		\includegraphics[scale=0.5]{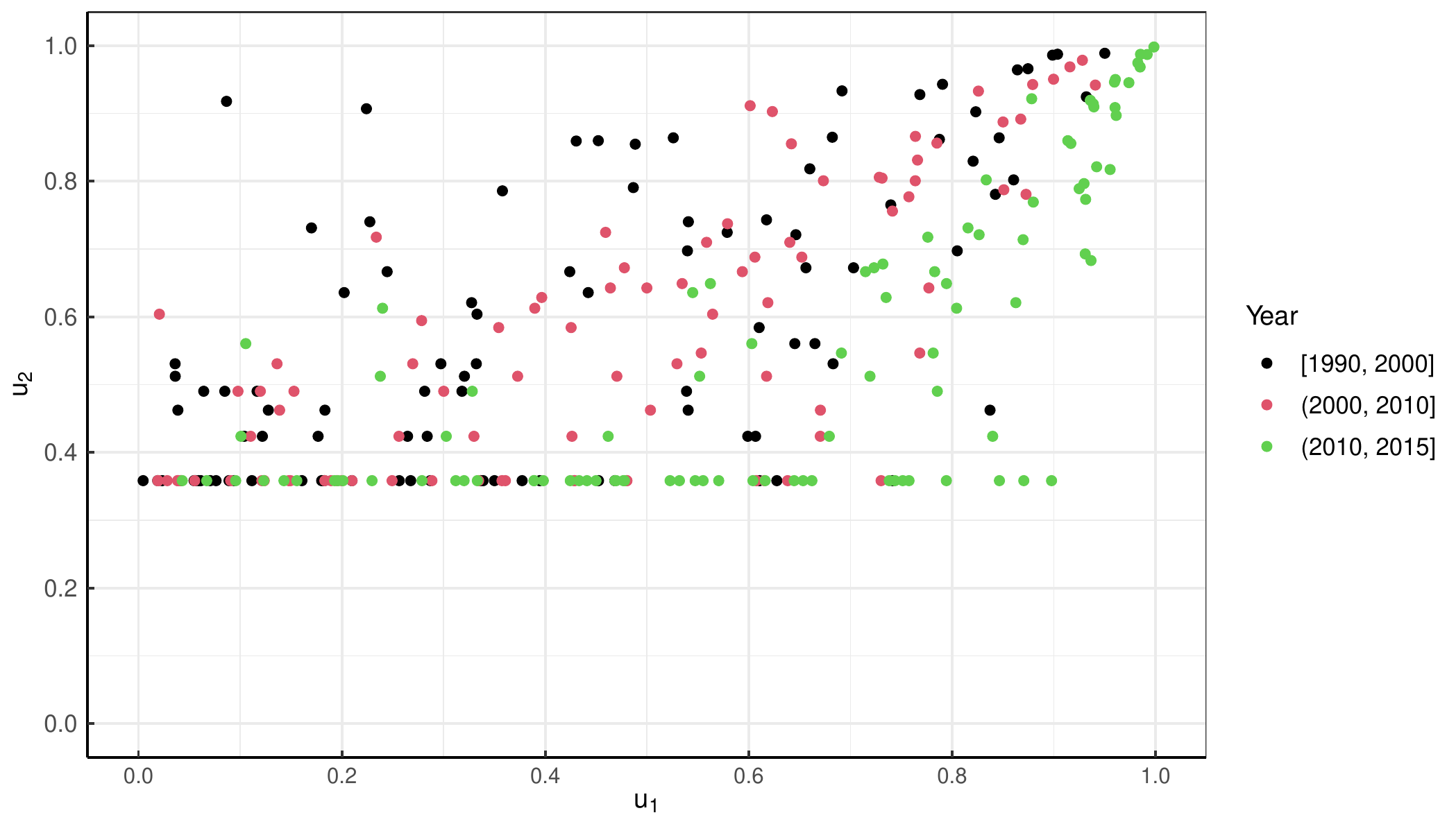} 
		\caption{
			Scatterplot of the uniform transformed data of economic losses and the number of casualties for different time intervals indicated by the color.
		}
		\label{fig:Losses-Casualties}
	\end{figure}

\vspace{0.3cm}	
	To understand the association, we create uniform transformed data $(u_1,u_2)$
	by applying the probability integral transformation to
	the economic losses and the number of casualties.
	 	The empirical Spearman's \textit{rho} and Kendall's \textit{tau} are 0.634 and 0.479 respectively.
Figure \ref{fig:Losses-Casualties} displays the pairwise dependence structures evaluated at $(u_1,u_2)$ as a function of the \textit{Year}, which indeed indicates changes in dependence over the years.

	In order to accommodate the asymmetric feature exhibited, we apply a mixed copula to model the dependence between the continuous and semi-continuous outcomes.
	The modelling of mixed copula is also discuss in \cite{shi2018pair}
	when the outcome follows a semi-continuous distribution.
	The idea is easily extended to the general mixed case.
	Here we also refer to the most recent work of \cite{chang2019prediction}, which proposes a copula regression to handle mixed continuous and discrete response variables.
	
	For a portfolio of $n$ observations $(y_{i1},y_{i2}; \; i=1,\ldots,n)$, the joint density function of $(Y_1,Y_2)$ can be written as
	\begin{equation}
	\label{joint-density-mixed-copula}
	f_{Y_{1},Y_2}(y_{i1},y_{i2})=\begin{cases}
	f_{Y_1}(y_{i1})\left[
	h_{2|1}(F_{Y_{1}}(y_{i1}),F_{Y_{2}}(y_{i2})) - h_{2|1}(F_{Y_{1}}(y_{i1}),F_{Y_{2}}(y_{i2}-1))
	\right], & y_{i2}\le u,\\
	f_{Y_1}(y_{i1})f_{Y_2}(y_{i2})c(F_{Y_{1}}(y_{i1}), F_{Y_{2}}(y_{i2})), & y_{i2} > u,
	\end{cases}
	\end{equation}
	where the density $f_{Y_j}(\cdot)$ and cdf $F_{Y_j}(\cdot)$ of the marginal distributions   ($i=1,2$) are specified by \eqref{pdf:LMGA}, \eqref{cdf:GLMA} and \eqref{eq:f2}, \eqref{eq:F2} respectively. Here
	$h_{2|1}(u_1, u_2)=\partial C(u_1,u_2)/\partial u_1$ is the $h$-function of bivariate copula.
	The log-likelihood function with copula parameters $\bm{\gamma}$ and marginal parameters $\bm{\xi}$ is given by
	\begin{align}
	\label{log-likelihood-mixed-copula}
	\ell(\bm{\gamma},\bm{\xi})= \sum_{i=1}^{n}\log f_{Y_1, Y_2}(y_{i1},y_{i2}).
	\end{align}
		Here we use the inference functions of margins (IFM) method proposed discussed in \cite{joe1997multivariate} and \cite{nelsen2007introduction} to reduce the computational burden.
	Compared with the standard ML method which is a full likelihood approach estimating all parameters simultaneously, the IFM method is a two-step approach. In the first step the marginals are fitted independently to obtain the estimates of the parameters $\bm{\xi}$.
	In the second step the joint log-likelihood of the mixed copula given in \eqref{log-likelihood-mixed-copula} is maximized over
	the copula-related parameters fixed as estimated in the first step of the method.

	
	Table \ref{tab:estimates - copula} summarizes the
	copula parameter estimates, together with the log-likelihood and
	information statistics of the survival MGL and the survival MGL-EV copula, along with four other copula candidates.  
		\begin{itemize}
		\item For the survival MGL, $\bm{\gamma}=\delta$,
		\item  For the survival MGL-EV, $\bm{\gamma}=\delta$,
		\item
		for the MGB2
		copula, $\bm{\gamma}$ is a 3-vector of $(p_1, p_2, q)^T$,
		\item for the Gumbel copula,
		$\bm{\gamma}=\delta$,
		\item for the Gaussian copula, $\bm{\gamma}=\rho$,
		\item
		for the Student $t$ copula, $\bm{\gamma}= (\rho, v)^T$ and the degree of freedom $v$ is determined by ML.
	\end{itemize}	
	In terms of the BIC, the survival MGL copula is preferred.
In order to assess the quality of the copula fit in the tails, we focus on the model fit in the upper tails.
The model fit errors $e_{A}(C)$ in upper regions, e.g., $A=\left[0.95,1\right]$ and $A=\left[0.99,1\right]$, are reported in Table \ref{tab:estimates - copula}.
One can see that although survival MGL-EV and Gumbel copula might be more appropriate for modelling the data in $\left[0.95,1\right]$,
the survival MGL copula has a best performance in the region  $\left[0.99,1\right]$.

	\begin{table}[htbp]
		\centering
		\caption{Estimates and goodness fit of candidate of copulas for the Chinese earthquake loss data set.}
				\setlength{\tabcolsep}{1.3mm}{
			\begin{threeparttable}
				\begin{tabular}{lcccccccc}
			\toprule
			\multirow{2}[0]{*}{Copula} & \multicolumn{3}{c}{\multirow{2}[0]{*}{Estmates}} & \multirow{2}[0]{*}{Loglike} & \multirow{2}[0]{*}{AIC} & \multirow{2}[0]{*}{BIC} & \multirow{2}[0]{*}{$e_{\left[0.95,1\right]^2}$} &\multirow{2}[0]{*}{$e_{\left[0.99,1\right]^2}$}\\
			\\
			\hline
			\multirow{2}[0]{*}{Gaussian} & $\hat{\rho}=0.698$ &    -   &    -   & \multirow{2}[0]{*}{-3022.35} & \multirow{2}[0]{*}{6046.69} & \multirow{2}[0]{*}{6050.36}& \multirow{2}[0]{*}{66.85}  & \multirow{2}[0]{*}{0.29} \\
			& (0.028) &  -     &   -    &       &       &  \\
			\multirow{2}[0]{*}{Student $t$} & $\hat{\rho}=0.691$ & $\hat{v}=4.450$ &  -     & \multirow{2}[0]{*}{-3018.91} & \multirow{2}[0]{*}{6041.82} & \multirow{2}[0]{*}{6049.17} & \multirow{2}[0]{*}{57.44} & \multirow{2}[0]{*}{0.21}
			\\
			& (0.036) & (2.096) &    -   &       &       &  \\
			\multirow{2}[0]{*}{Gumbel} &$\hat{\alpha}= 1.922$ &    -   &   -    & \multirow{2}[0]{*}{-3009.92} & \multirow{2}[0]{*}{6021.84} & \multirow{2}[0]{*}{6025.51}&  \multirow{2}[0]{*}{{1.71}}&  \multirow{2}[0]{*}{0.08}
			 \\
			& (0.096) &   -    &  -     &       &       &  \\
			\multirow{2}[0]{*}{Survival MGL} & $\hat{\delta}=2.763$ &  -     &  -     & \multirow{2}[0]{*}{{-3009.46}} & \multirow{2}[0]{*}{{6020.92}} & \multirow{2}[0]{*}{{\textbf{6024.59}}} 
			& \multirow{2}[0]{*}{{2.66}}& \multirow{2}[0]{*}{\textbf{0.06}}
			\\
			& (0.247) &  -     &  -     &       &       &  \\
								\multirow{2}[0]{*}{Survival MGL-EV} & $\hat{\delta}=1.861$ & & & \multirow{2}[0]{*}{{	-3009.88}} & \multirow{2}[0]{*}{{6021.75}} & \multirow{2}[0]{*}{{{6025.42}}} &\multirow{2}[0]{*}{\textbf{1.71}} & 
								\multirow{2}[0]{*}{0.08}
								\\
				&(0.160)  & & &       &       &  \\
			\multirow{2}[0]{*}{MGB2} & $\hat{p_1}=1.485$ & $\hat{p_2}=1.322$ & $\hat{q}=0.816	$ & \multirow{2}[0]{*}{\textbf{-3007.30}} & \multirow{2}[0]{*}{\textbf{6020.60}} & \multirow{2}[0]{*}{{6031.62}} 
			&\multirow{2}[0]{*}{1.97}&\multirow{2}[0]{*}{0.07}
			\\
			& (2.347) & (0.785) & (0.530) &       &       &  \\
			\bottomrule
		\end{tabular}%
		\begin{tablenotes}
		\footnotesize
		\item Notes: The square fit error $e_{A}(C)$ in regions $A = \left[0.95,1\right]^2$ and $A = \left[0.99,1\right]^2$ are resealed by $\times 10^8$.
		The standard error is reported in brackets.
	\end{tablenotes}
	\end{threeparttable}
}
		\label{tab:estimates - copula}%
	\end{table}%

The introduction of  covariates into the dependence parameter can improve the model fitting. 
	We introduce here the covariate \textit{{Year}} into the dependence parameter in the survival MGL and survival MGL-EV regression model.
	The natural cubic splines are used to allow flexible relationships between \textit{{Year}} and the dependence parameter.
	We use the log link function to obtain real value number $\log \delta$, leading to the model
	\begin{equation*}
		\log \delta_{i}=\text{ns}_i(\text{Year})=\beta_1b_1(\text{Year})+...+\beta_kb_k(\text{Year}),
	\end{equation*}
	where $\text{ns}_i(x)$ denote  natural cubic splines with spline basis $b_1(x),...,b_p(x)$. 
	The 33.3\% and 66.7\% percentiles of \textit{Year} are used as two knots for the natural cubic spline, and there are four coefficients $\beta_{k}, k = 1,...,4$ to be estimated in the copula regression.  
	The IFM method is used to estimate the regression coefficients.
	
	{Table} \ref{tab:estimates - earth} reports the estimates and standard errors of regression coefficients, together with the log-likelihood and information statistics for the survival MGL and survival MGL-EV copula regression. The estimation results are also reported for  a corresponding Gumbel regression model.
	In terms of the AIC and BIC value the survival MGL copula regression provides a better overall fit.
	 
\begin{figure}[htbp]
		\centering
		\includegraphics[scale=0.55]{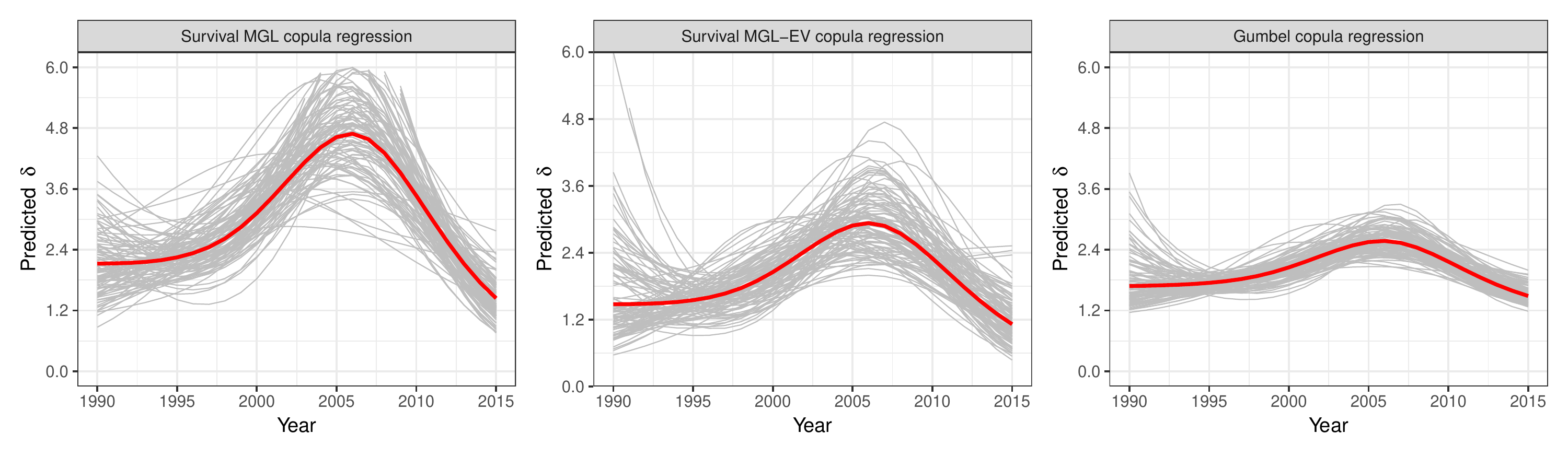}
		\caption{The predicted value of copula parameter with different value of covariate \textit{Year} for the Chinese earthquake loss data set.
			The gray lines are generated based on the simulated coefficients for the natural cubic spline, and their values are generated by a multivariate normal distribution with  mean equal to the  ML estimates  of $\beta_k$ for $k=1,...,4$, and the covariance matrix being the inverse of the Hessian matrix.}
		\label{fig-earth-predicted}
	\end{figure}

	Figure \ref{fig-earth-predicted} presents the relationship between \textit{Year} and the dependence parameter.
	One can see that the non-linear relationship appear in three cases, which displays the tendency of dependence rising up at the beginning and declining later on.
	The magnitude 8.0 Sichuan earthquake from 2008 was the strongest earthquake 
		in China in over 50 years.  After 2008 the
		Chinese government has conducted many disaster management-related projects, responding to the need for disaster prevention and mitigation by integrating livelihood assistance, disaster risk reduction, sharing knowledge and practice, technical support, capacity building, and policy advocacy after 2008, resulting in a  declining trend of the dependence.

		\begin{table}[htbp]
		\centering
		\caption{Estimates and goodness fit for survival MGL and Gumbel regression models for the Chinese earthquake loss data set.}
		\setlength{\tabcolsep}{5.5mm}{
			\begin{threeparttable}
				\begin{tabular}{c|cccccc}
					\toprule
					\multicolumn{1}{c|}{\multirow{2}[0]{*}{Parameters	}} 	 & \multicolumn{2}{c}{Survival MGL} & 
					\multicolumn{2}{c}{Survival MGL-EV}  &\multicolumn{2}{c}{Gumbel} \\
		& Estimates & S.E.   & Estimates & S.E. 	& Estimates & S.E. \\
\hline
 $\beta_1$ &   0.634	&      0.333	 & 0.333 &0.389&   -0.250 &      0.495 \\
$\beta_2$&1.722 &      0.292	 &   1.327 &0.311&  1.087 &      0.366 \\
$\beta_3$&1.846 &      0.386	 &   1.051	 &0.355&-0.608 &      0.470 \\
$\beta_4$&-0.250 &      0.341	 &   -0.211 &0.357&  -0.422 &      0.480 \\
					\hline
					Loglike & \multicolumn{2}{c}{\textbf{-3002.22}} & \multicolumn{2}{c}{-3004.76}& \multicolumn{2}{c}{-3004.89} \\
					AIC   & \multicolumn{2}{c}{\textbf{6012.43}} & \multicolumn{2}{c}{6017.52}& \multicolumn{2}{c}{6017.78} \\
					BIC   & \multicolumn{2}{c}{\textbf{6027.12}} & \multicolumn{2}{c}{6032.22}& \multicolumn{2}{c}{6032.48} \\
					\bottomrule
				\end{tabular}%
				\begin{tablenotes}
					\footnotesize
					\item Notes: In order to avoid boundary problem in MLE procedures, 
					we consider a log link function obtaining real values for Gumbel copula regression:
					$\log(\delta_{i}-1)=\text{ns}_i(\text{Year})$ for all $\delta_i>1$.
				\end{tablenotes}
			\end{threeparttable}
		}
		\label{tab:estimates - earth}%
	\end{table}%

	\newpage
	\section{Summary and concluding remarks}\label{section:conclusion}
	The main proposals in this paper are the MGL copula model for accommodating  non-elliptical and
	asymmetric dependence structures, and dynamic dependence modelling using  corresponding copula regression models. Based on the heavy-tailedness
	from the univariate GLMGA distribution, this new copula class and its survival version can
	capture positive lower/upper tail dependence.
	The proposed copula
	features asymmetric relationships using  only one dependence parameter and demonstrates flexibility for modelling multi-dimensional asymmetry.
	The probabilistic characteristics of the proposed copula are discussed and the corresponding extreme-value copula is obtained.
	The proposed copula model is effective in regression modelling of the dependence structure using covariate information.
	ML estimation can be quite easily performed as the joint pdf is given in closed form even in  high dimensions. 
	We also
	implement the proposed method in a user-friendly R package:
\texttt{rMGLReg} that can provide a nice visualization tool for interpreting the proposed copula and 
serve as a convenient tool for actuarial practitioners to investigate the nonlinear dynamic dependence pattern.

	Considering  bivariate copulas as building blocks for many multivariate dependence models using for instance vine copulas, the potential of the proposed copula for building up more complex multivariate dependence models should be the subject of future research. There are ongoing studies on its potential in modelling real datasets that have more dimensions and more complex dependence structure.


		\section*{Supplementary Material}
	\begin{description}
		\item[R-package for github routine:] R package: \texttt{rMGLReg} containing code to display the properties of the proposed models and 
		perform the estimation methods described in the paper. 
		The package also contains all datasets used as examples in the paper. The package can be found at 
		\url{https://github.com/lizhengxiao/rMGLReg} for more details.
	\end{description}

	\appendix
	\section*{Appendices}
	\addcontentsline{toc}{section}{Appendices}
	\renewcommand{\thesubsection}{\Alph{subsection}}
	\subsection{Proof of Proposition 2.1 }
	\label{app:pairwise}
	From the model specification we   obtain the following properties:
	\[
	\mathbb{E}(Y_j|\Theta)=\left(\frac{\Theta}{2b_j}\right)^{\sigma_j}\frac{\Gamma(\frac{1}{2}-\sigma_j)}{\Gamma(\frac{1}{2})}
	\]
	and
	\[
	\text{Cov}(Y_j,Y_{j'}|\Theta)=\begin{cases}
	\left(\frac{\Theta}{2b_j}\right)^{2\sigma_j}\frac{\Gamma(\frac{1}{2})\Gamma(\frac{1}{2}-2\sigma_j)-\Gamma^2(\frac{1}{2}-\sigma_j)}{\Gamma^2(\frac{1}{2})}& \text{if}\quad j= j'\\
	0& \text{if} \quad j\ne j',
	\end{cases}
	\]
	and
	\[
	\mathbb{E}(\Theta^r)=\frac{\Gamma(r+a)}{\Gamma(a)}
	\quad \text{and} \quad
	\mathbb{E}(\Theta)=a
	\quad \text{and} \quad
	\text{Var}(\Theta)=a.
	\]
	These properties lead us to the unconditional mean of the univariate MGL distribution (also known as GLMGA distribution):
	\begin{align*}
	\mathbb{E}(Y_j)&=\mathbb{E}\left[\mathbb{E}(Y_j|\Theta)\right]
	=(2b_j)^{-\sigma_j}\frac{B(\frac{1}{2}-\sigma_j, a+\sigma_j)}{B(\frac{1}{2},a)},\\
	\text{Var}(Y_j)
	&=\mathbb{E}\left[\text{Var}(Y_j|\Theta)\right]+\text{Var}\left[\mathbb{E}(Y_j|\Theta)\right]
	=\left[\mathbb{E}(Y_j)\right]^2\left[
	\frac{B(a+2\sigma_j,a)B(\frac{1}{2}-2\sigma_j,\frac{1}{2})}{B(a+\sigma_j,a+\sigma_j)B(\frac{1}{2}-\sigma_j,\frac{1}{2}-\sigma_j)}
	-1
	\right].
	\end{align*}
	It is straightforward to see that in the case where $j\ne j'$ we have,
	\begin{align*}
	\text{Cov}(Y_j,Y_{j'})&=\mathbb{E}\left[\text{Cov}(Y_j,Y_{j'}|\Theta)\right]
	+\text{Cov}\left[\mathbb{E}(Y_j|\Theta),\mathbb{E}(Y_{j'}|\Theta)
	\right],\\
	&=\text{Cov}\left[\left(\frac{\Theta}{2b_j}\right)^{\sigma_j }\frac{\Gamma(\frac{1}{2}-\sigma_j)}{\Gamma(\frac{1}{2})},
	\left(\frac{\Theta}{2b_{j'}}\right)^{\sigma_{j'} }\frac{\Gamma(\frac{1}{2}-\sigma_{j'})}{\Gamma(\frac{1}{2})}
	\right]\\
	&=(2b_j)^{-\sigma_j}(2b_{j'})^{-\sigma_{j'} }\frac{\Gamma(\frac{1}{2}-\sigma_j)\Gamma(\frac{1}{2}-\sigma_{j'})}{\Gamma^2(\frac{1}{2})}
	\text{Cov}\left[\Theta^{\sigma_j},\Theta^{\sigma_{j'}}
	\right]\\
	&=\mathbb{E}(Y_j)\mathbb{E}(Y_{j'})\left[
	\frac{B(a+\sigma_j+\sigma_{j'},a)}{B(a+\sigma_j,a+\sigma_{j'})}
	-1\right].
	\end{align*}
	For $j\ne j'$ we then obtain
	\begin{eqnarray*}
	\mbox{Corr}(Y_{j},Y_{j'})
	&=&
	   \frac{\mbox{Cov}(Y_{j},Y_{j'})}{\sqrt{\mbox{Var}(Y_j)       \mbox{Var}(Y_{j¡¯})}} \\
	&=&\frac{\frac{B(a+\sigma_j+\sigma_{j'},a)}{B(a+\sigma_j,a+\sigma_{j'})}
		-1}{
		\sqrt{\left[\frac{B(a+2\sigma_j,a)B(\frac{1}{2}-2\sigma_j,\frac{1}{2})}{B(a+\sigma_j,a+\sigma_j)B(\frac{1}{2}-\sigma_j,\frac{1}{2}-\sigma_j)}
			-1\right]\left[\frac{B(a+2\sigma_{j'},a)B(\frac{1}{2}-2\sigma_{j'},\frac{1}{2})}{B(a+\sigma_{j'},a+\sigma_{j'})B(\frac{1}{2}-\sigma_{j'},\frac{1}{2}-\sigma_{j'})}
			-1\right]}}
	\end{eqnarray*}
	Note that if $\sigma_j=\sigma_{j'}:=\sigma$, \eqref{pairwise correlation}  simplifies to
	\[
	\text{Corr}(Y_j,Y_{j'})=\frac{\frac{B(a+2\sigma,a)}{B(a+\sigma,a+\sigma)}-1}{\frac{B(a+2\sigma,a)}{B(a+\sigma,a+\sigma)}\frac{B(\frac{1}{2}-2\sigma,\frac{1}{2})}{B(\frac{1}{2}-\sigma,\frac{1}{2}-\sigma)}-1}.
	\]
	Since for any $m$, $\lim_{a\to+\infty}\frac{\Gamma(a+m)}{\Gamma(a)a^m}=1$, we have that for any fixed $\sigma$,
	\begin{align*}
	\lim_{a\to+\infty}{\frac{B(a+\sigma_j+\sigma_{j'},a)}{B(a+\sigma_j,a+\sigma_{j'})}
		-1}&=
	\lim_{a\to+\infty}{\frac{\Gamma(a+\sigma_j+\sigma_{j'})\Gamma(a)}{\Gamma(a+\sigma_j)\Gamma(a+\sigma_{j'})}-1}=0.
	\end{align*}
		This implies $\lim_{a\to\infty}\text{Corr}(Y_j,Y_{j'})=0$ if $j\ne j'$.

\subsection{Proof of Proposition \ref{prop: simulation}}\label{app:simulation}
	Random samples $(Y_1,...,Y_d)$ from the $\text{MGL}(\bm{\sigma},a,\bm{b})$ distribution can be simulated  using the conditional distribution given in Proposition \ref{prop: conditional}  through the following steps:
	\begin{itemize}
		\item $Y_1$ is generated using the quantile function  of the GLMGA$(\sigma_1, a, b_1)$ distribution:
		\[
		Y_1 = (2b_1)^{-\sigma_1}\left[\frac{I^{-1}_{\frac{1}{2},a}(1-U_1)}{1-I^{-1}_{\frac{1}{2},a}(1-U_1)}\right]^{-\sigma_1}
		= (2b_1M_1)^{-\sigma_1} .
		\]
		\item $Y_2$ is generated using the quantile function of the  GLMGA$(\sigma_2, a_2, b_2^*)$ distribution 
		with $a_2 = a + \frac{1}{2}$ and $b_2^*=b_2\left[1+Y_1^{-1/\sigma_1}/(2b_1)\right]$:
		\[
		Y_2 = (2b_2^*)^{-\sigma_2}\left[\frac{I^{-1}_{\frac{1}{2},a_2}(1-U_2)}{1-I^{-1}_{\frac{1}{2},a_2}(1-U_2)}\right]^{-\sigma_2}
		= (2b_2^*Z_2)^{-\sigma_2}= \left[2b_2M_2\right]^{-\sigma_2}.
		\]
		
		\item ......
		\item $Y_d$ is generated using the quantile function of the GLMGA$(\sigma_d, a_d, b_d^*)$ distribution with $a_d = a + \frac{d-1}{2}$ and $b^{*}_d=b_d \left[1+\sum_{j=1}^{d-1}Y_j^{-1/\sigma_j}/(2b_j)\right]$:
		\[
		Y_d = (2b_d^*)^{-\sigma_d}\left[\frac{I^{-1}_{\frac{1}{2},a_d}(1-U_d)}{1-I^{-1}_{\frac{1}{2},a_d}(1-U_d)}\right]^{-\sigma_d}
		= (2b_d^*Z_d)^{-\sigma_d}= \left[2b_dM_d\right]^{-\sigma_d}.
		\]

		Finally, the random samples $(U_1^*,...,U_d^*)$ from $C^{MGL}(\cdot ;a)$ can  be obtained by
		\[
		U_j^*=F(Y_j;\sigma_j,a, b_j)=1-I_{{\frac{1}{2}}, a}\left(\frac{M_j}{1+M_j}\right),	\quad		\text{for} \quad j=1,...,d, 
		\]
		where $F(;\sigma_j,a, b_j)$ is the cdf of the univariate GLMGA distribution given in \eqref{cdf:GLMA}.
		The random samples of $C^{MGL}(\cdot;\delta)$ are generated substituting $\delta=1/a$.
	\end{itemize}

	\subsection{Proof of Proposition \ref{prop:tail-dependence}}
	We first define $Y_1=F_1^{-1}(U_1)$ and $Y_2=F_2^{-1}(U_2)$, where $U_1, U_2$ are independent uniformly (0,1) distributed and 
	$F_j$ ($j=1,2$) represent the cdf of the GLMGA distribution with
	parameters $(\sigma_j, a, b_j)$ respectively. Using Proposition \ref{prop: conditional} the vector $(Y_1,Y_2)$ satisfies
	\begin{align*}
	&Y_1|Y_2=y_2 \sim \text{GLMGA}(\sigma_1, a+\frac{1}{2}, b_1(1+\frac{y_2^{-1/\sigma_2}}{2b_2})) \\
	&Y_2|Y_1=y_1 \sim \text{GLMGA}(\sigma_2, a+\frac{1}{2}, b_2(1+\frac{y_1^{-1/\sigma_1}}{2b_1}))
	\end{align*}
	For the upper tail dependence index, we use
	\begin{align*}
	\lambda_u&=\lim_{u\to 1^-}\Pr\left[
	Y_1>F_1^{-1}(u)|Y_2=F_2^{-1}(u)
	\right] +\lim_{u\to 1^-}\Pr\left[
	Y_2>F_2^{-1}(u)|Y_1=F_1^{-1}(u)
	\right] \\
	&=
	\lim_{u\to 1^-}
	I_{\frac{1}{2},a+\frac{1}{2}}\left(
	\frac{I^{-1}_{\frac{1}{2}, a + \frac{1}{2}}(1-u)}{1 + I^{-1}_{\frac{1}{2}, a + \frac{1}{2}}(1-u)}
	\right) +  \lim_{u\to 1^-}
	I_{\frac{1}{2},a+\frac{1}{2}}\left(
	\frac{I^{-1}_{\frac{1}{2}, a + \frac{1}{2}}(1-u)}{1 + I^{-1}_{\frac{1}{2}, a + \frac{1}{2}}(1-u)}
	\right)
	\\
	&=0.
	\end{align*}
	For the lower tail dependence index, we use
	\begin{align*}
	\lambda_l&=\lim_{u\to 0^+}\Pr\left[
	Y_1 \le F_1^{-1}(u)|Y_2=F_2^{-1}(u)
	\right] + \lim_{u\to 0^+}\Pr\left[
	Y_2 \le F_2^{-1}(u)|Y_1=F_1^{-1}(u)
	\right]\\
	&=2 - 2\lim_{u\to 0^+}
	I_{\frac{1}{2},a+\frac{1}{2}}\left(
	\frac{I^{-1}_{\frac{1}{2}, a + \frac{1}{2}}(1-u)}{1 + I^{-1}_{\frac{1}{2}, a + \frac{1}{2}}(1-u)}
	\right)
	\\
	&=2-2I_{\frac{1}{2},a+\frac{1}{2}}\left(\frac{1}{2}\right)\\
	&=2-2I_{\frac{1}{2},\frac{1}{\delta}+\frac{1}{2}}\left(\frac{1}{2}\right).
	\end{align*}

	\subsection{Domain of attraction and extreme-value copula}\label{App: EV-copula}
	For the proof of Proposition \ref{prop:MGL-EV} we need the following intermediate result concerning the regular variation of the function $t(\cdot;a)$.	
	\begin{lemma}
				The function $t(u; a)=\frac{I^{-1}_{\frac{1}{2},{a}}(1-u)}{1-I^{-1}_{\frac{1}{2},{a}}(1-u)}$ is  regularly varying at the origin with index $-1/a$:
		\begin{equation}
		\lim_{s \to 0}\frac{t(su_j; a)}{t(s; a)} = u^{-1/a}, \quad u>0.
		\label{lemma:q-limit}
		\end{equation}
	\end{lemma}
	
	\noindent
{\bf Proof.} Since the cdf $F$ of a GLMGA distribution is regularly varying near 0 with index $a/\sigma$, its inverse, the  quantile function given in  \eqref{qf:GLMA}, is also regularly varying at 0 with index $\sigma/a$. Since $t$ is proportional to the $-1/\sigma$ power of the quantile function  the result follows. $\Box$\\

\noindent
{\bf Proof of Proposition \ref{prop:MGL-EV}}
				Clearly, the boundary values of $\ell$ are given by $\ell(u_1, 0)=u_1$, $\ell(0, u_2)=u_2$ and $\ell(0,0)=0$.	
		Moreover
		\begin{align*}
		\lim_{s\to 0^+}\frac{1-\bar{C}^{MGL}(1-su_1, 1-su_2;\delta)}{s}&=\nonumber
		\lim_{s\to 0^+}\frac{su_1 + su_2 - C^{MGL}(su_1, su_2;\delta)}{s} \nonumber\\
		&= u_1 + u_2\nonumber \\ 
		&\quad - \lim_{s\to 0^+} u_1\Pr\left(U_2\le su_2| U_1=su_1\right)\nonumber\\
		&\quad  - 	\lim_{s\to 0^+} u_2\Pr\left(U_1\le su_1| U_2=su_2\right).
		\label{eq: ev-cond}
		\end{align*}
		Let $Y_1=F^{-1}_{1}(U_1)$ and $Y_2=F^{-1}_{2}(U_2)$ where $F_j$ is the distribution function of the univariate GLMGA distribution with parameters $(\sigma_j,a,b_j)$, $j=1,2$. 
		The conditional	probability function can be evaluated using Proposition \ref{prop: conditional} and is given by
		\begin{align*}
		\Pr\left(U_2 \le su_2| U_1=su_1\right) &= \Pr\left(Y_2\le F^{-1}_{2}(su_2)| Y_{1} = F^{-1}_{1}(su_1)\right) \nonumber\\
		&=1-I_{\frac{1}{2}, a+\frac{1}{2}}\left[\frac{t(su_2;a)}{t(su_1;a)+ t(su_2;a)+ 1} \right] \nonumber\\
		& = 1-I_{\frac{1}{2}, a+\frac{1}{2}}\left[\frac{\frac{t(su_2;a)}{t(s;a)}}{\frac{t(su_2;a)}{t(s;a)} + \frac{ t(su_1;a)}{t(s;a)}+ \frac{1}{t(s;a)}} \right].
		\end{align*}
		A similar expression holds for $\Pr\left(U_1 \le su_1| U_2=su_2\right)$. Since $1/t(s;a) \to 0^+$ as $s\to 0^+$ and the only remaining terms depending on $s$ are $\frac{t(su_1;a)}{t(s;a)}$ and $\frac{t(su_2;a)}{t(s;a)}$, the limit can be obtained using Lemma \ref{lemma:q-limit}:
		\begin{align*}
		\ell(u_1,u_2)&=\lim_{s\to 0^+}\frac{1-\bar{C}^{MGL}(1-su_1, 1-su_2;\delta)}{s}\nonumber\\
		&=
		u_1I_{\frac{1}{2}, a+\frac{1}{2}}\left(\frac{u_2^{-1/a}}{u_1^{-1/a}+u_2^{-1/a}}\right)
		+u_2 I_{\frac{1}{2}, a+\frac{1}{2}}\left(\frac{u_1^{-1/a}}{u_1^{-1/a}+u_2^{-1/a}}\right),\nonumber\\
		& = 	u_1I_{\frac{1}{2}, \frac{1}{\delta}+\frac{1}{2}}\left(\frac{u_2^{-\delta}}{u_1^{-\delta}+u_2^{-\delta}}\right)
		+u_2 I_{\frac{1}{2}, \frac{1}{\delta}+\frac{1}{2}}\left(\frac{u_1^{-\delta}}{u_1^{-\delta}+u_2^{-\delta}}\right),
		\end{align*}
		where $\delta = 1/a$.		
		
		Setting $A_{\delta}(w)=\ell(w, 1-w)$,   $w \in [0,1]$, we obtain the form given in \eqref{eq:survival MGL-EV1} and \eqref{eq:Pickands}.
		The resulting copula is an extreme value copula since $A_{\delta}(w)=\ell(w, 1-w)$ defined by \eqref{eq:Pickands} is a convex function 
		satisfying $\max(1-w,w)\le A_{\delta}(w)\le 1$ for $0\le w\le 1$. The lower bound, $A_{\delta}(w)=\max(1-w,w)$, corresponds to complete dependence, whereas the upper bound, $A_{\delta}(w)=1$, corresponds to independence. $\Box$\\
		
		\noindent
\textbf{The $d$-dimensional extreme value copula} The extreme value copula $\bar{C}^{\text{MGL-EV}}$ of the survival MGL copula if given by
		\[
		\bar{C}^{\text{MGL-EV}}(u_1,..., u_d;\delta)= \exp\left[-\ell(-\log u_1,...,-\log u_d)\right],
		\]
		where the stable tail dependence function $\ell:\left[0,\infty \right.\left.\right) ^d \to \left[0,\infty\right.\left.\right)$ is given by
		\[
		\ell(u_1,...,u_d)=\sum_{j=1}^{d}u_jI_{\frac{1}{2}, \delta+\frac{1}{2}}\left(1-\frac{u_j^{-1/a}}{\sum_{j=1}^{d}u_j^{-1/a}}\right).
		\]
		The density of the $d$-dimensional extreme-value copula $\bar{C}^{\text{MGL-EV}}$ is of the form
		\[
		\bar{c}^{\text{MGL-EV}}(u_1,..., u_d;\delta)=
		\frac{\bar{C}^{\text{MGL-EV}}(u_1, ..., u_d;\delta)}{\prod_{j=1}^{d}u_j}\sum_{m=1}^{d}(-1)^{d-m}\sum_{\pi:\abs{\pi}=m}\prod_{B\in\pi}D_B\ell(z_1,...,z_d)|_{z_1=-\log u_1,...,z_d = -\log u_d}.
		\]


	\subsection{The gradient of the log-likelihood for survival MGL copula}\label{App: gradient}
	We obtain derivatives of the log-likelihood \eqref{eq: log-likelihood-MGL} in Section \ref{section: regression} with respect to model parameters.
	With $\delta_i=\exp(\bm{x}_i^T\bm{\beta})$, 
	$\phi'(x):=\frac{\partial \log \Gamma(x)}{\partial x}|_{x=x}$, $\phi''(x):=\frac{\partial^2 \log \Gamma(x)}{\partial x^2}|_{x=x}$,
	$m'(x,\frac{1}{\delta_i})=:
	{\frac{\partial I^{-1}_{\frac{1}{2},z}(x)}{\partial z}}{|z=\frac{1}{\delta_i}}$ 
	and 
	$m''(x,\frac{1}{\delta_i})=:
	\frac{\partial^2 I^{-1}_{\frac{1}{2},z}(x)}{\partial z^2}|z=\frac{1}{\delta_i}$, 
	the first-order derivatives are given by
	
		\begin{align}
	\frac{\partial \ell({\bm{u}}_{1}, ..., {\bm{u}}_{d};\bm{\beta})}{\partial \beta_{h}} &= 
	-\frac{x_{ij}}{\delta_{i}}\left\{
	(d-1)\sum_{i=1}^{n}\phi'\left(\frac{1}{\delta_i}\right) 
	+ \sum_{i=1}^{n}\phi'\left(\frac{1}{\delta_i} + \frac{d}{2}\right)
	-  d\sum_{i=1}^{n}\phi'\left(\frac{1}{\delta_i}+\frac{1}{2}\right) \right. \nonumber \\
	&\quad \quad + 
	\sum_{i=1}^{n}\sum_{j=1}^{d} \log \frac{I^{-1}_{\frac{1}{2},\frac{1}{\delta_i}}({u}_{ij})}{1-I^{-1}_{\frac{1}{2},\frac{1}{\delta_i}}({u}_{ij})} \nonumber \\
&\quad \quad	+\sum_{i=1}^n\left(\frac{1}{\delta_i}+\frac{1}{2}\right)\sum_{j=1}^d\frac{m'(u_{ij}, \frac{1}{\delta_i})}{I^{-1}_{\frac{1}{2},\frac{1}{\delta_i}}({u}_{ij})}
	+\sum_{i=1}^n\left(\frac{1}{\delta_i}+\frac{1}{2}\right)\sum_{j=1}^d\frac{m'(u_{ij}, \frac{1}{\delta_i})}{1-I^{-1}_{\frac{1}{2},\frac{1}{\delta_i}}({u}_{ij})}\nonumber \\
&\quad \quad + 
\sum_{i=1}^{n}\log\left(\sum_{j=1}^{d} \frac{I^{-1}_{\frac{1}{2},\frac{1}{\delta_i}}({u}_{ij})}{1-I^{-1}_{\frac{1}{2},\frac{1}{\delta_i}}({u}_{ij})} + 1\right)\nonumber \\
&\quad \quad \left.	+ \sum_{i=1}^n\left(\frac{1}{\delta_i}+\frac{d}{2}\right)
\left[\sum_{j=1}^{d} \frac{I^{-1}_{\frac{1}{2},\frac{1}{\delta_i}}({u}_{ij})}{1-I^{-1}_{\frac{1}{2},\frac{1}{\delta_i}}({u}_{ij})}+1\right]^{-1}
\sum_{j=1}^{d}\frac{m'({u}_{ij},\frac{1}{\delta_i})}{\left[{1-I^{-1}_{\frac{1}{2},\frac{1}{\delta_i}}({u}_{ij})}\right]^2}
\right\},
\label{eq:first-order}
	\end{align}
	for $h=0,...k$.

	Equating \eqref{eq:first-order} to zero, the maximum likelihood (ML)
	estimator $\hat{\beta_{h}}$ of ${\beta_{h}}$ is obtained by using the function \texttt{MGL.reg} in R package:\texttt{ rMGLReg }
	to minimize the negative log-likelihood with a given gradient.  
	

	\subsection{Economic loss: marginal modelling}\label{App: GLMGA}
	
In Table \ref{tab:estimates - data I} we provide the estimates, log-likelihood values
	(LL), as well as the Akaike Information Criterion (AIC) and the Bayesian Information Criterion (BIC) values, defined respectively as $\text{AIC}=-2\ell+2p$ and $\text{BIC}=-2\ell+p\log n$ where $\ell$ denotes the log-likelihood value, $p$ the number of model parameters and $n$ the number of observations.
	We use the \texttt{{optim()}} function in R which uses the Nelder-Mead method.
	Parameters are estimated by the MLE and standard errors are calculated using the observed information matrix.
	It is clear from Table \ref{tab:estimates - data I} that the GLMGA provide a better fit than the other four models, as it has the highest log-likelihood  and smallest AIC and BIC value.

	We also provide goodness-of-fit measures and the bootstrap P-values for the corresponding goodness-of-fit tests. In Table \ref{tab:gof}  we consider the Kolmogorov-Smirnov (KS), Cram\'er-von Mises (CvM) and Anderson-Darling (AD) test statistics and corresponding P-values, choosing for the models with
	small values of the KS, CvM and AD test statistics, or large values of the corresponding P-values. The P-values are obtained using the bootstrap method as developed in \cite{calderin2016modeling}.
	Here again the GLMGA model is prevailing with a P-value above 0.7, which provides a strong evidence for the best fit.
	
\begin{figure}[htbp]
		\centering
		\includegraphics[scale=0.7]{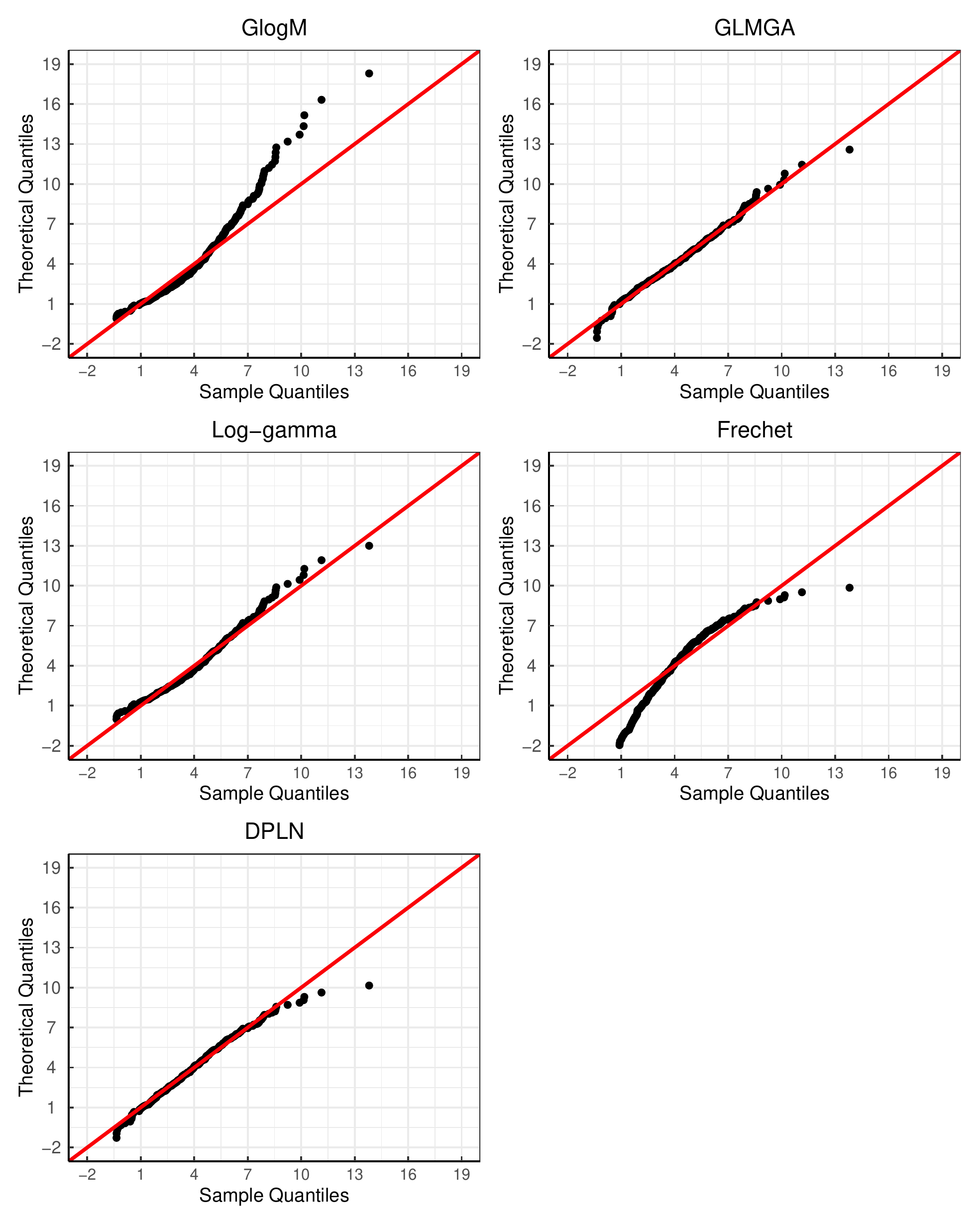} 
		\caption{Earthquake economic losses: QQ-plots of the log-transformed empirical  quantiles against the log-transformed estimated model quantiles.}
		\label{fig:QQ-plot-data I}
	\end{figure}

	In Figure \ref{fig:QQ-plot-data I} the QQ-plots  of the log-transformed empirical quantiles against the log-transformed estimated  quantiles of the 5 competing models are given. The correlation coefficients $R$ of these  QQ-plots are also given in Table \ref{tab:gof}: $R$  measures the degree of linearity in the QQ-plot and hence also the  goodness-of-fit with respect to the corresponding model.
	These QQ-plots also provide interesting information concerning the estimates of the VaR at extreme quantile levels. We can judge the appropriateness of the VaR estimates using the different competing models by comparing the model estimates of an extreme quantile $F^{-1}(p)$ with the quantile level $p$ close to $1-{1 \over n}$ with the empirical VaR, which is then close to the maximum value of the data set. In Table \ref{tab:VaR-competing models} we compare the empirical 95\%, 99\%, 99.8\% Var with the estimates of the model VaR  obtained from the different models.
	We report the relative deviations from the empirical $\text{VaR}$.
	Note that the Fr\'echet and DPLN models are lower than the empirical estimate, while the GLMGA model gives much more conservative estimates than GlogM and Log-gamma models.

	\begin{table}[tp]
		\centering
		\caption{Earthquake economic losses: model selection measures.}
		\begin{tabular*}{\hsize}{@{}@{\extracolsep{\fill}}lllllll@{}}
\toprule
			Distribution & \multicolumn{2}{l}{Estimates} &  \#Par. & LL    & AIC   & BIC   \\
			\hline
			\multirow{2}[0]{*}{GlogM} &$\hat{\sigma}$ & 1.426 (0.061) & \multirow{2}[0]{*}{2} & \multirow{2}[0]{*}{-1899.76} & \multirow{2}[0]{*}{3803.52} & \multirow{2}[0]{*}{3810.86}  \\
			& $\hat{\mu}$& 15.475 (2.042) &       &       &       &         \\
			\cmidrule{1-7}
			\multirow{3}[0]{*}{GLMGA} & $\hat{\sigma}$& 0.820 (0.074) & \multirow{3}[0]{*}{3} & \multirow{3}[0]{*}{\textbf{-1871.01}} & \multirow{3}[0]{*}{\textbf{3748.02}} & \multirow{3}[0]{*}{\textbf{3759.04}}  \\
			& $\hat{b}$& 0.005 (0.005) &       &       &       &         \\
			& $\hat{a}$ & 0.697 (0.153) &       &       &       &         \\
			\cmidrule{1-7}
			\multirow{2}[0]{*}{Log-gamma} & $\hat{\alpha}$ & 3.547 (0.280) & \multirow{2}[0]{*}{2} & \multirow{2}[0]{*}{-1878.35} & \multirow{2}[0]{*}{ 3760.70} & \multirow{2}[0]{*}{3768.05}  \\
			& $\hat{\beta}$ & 1.215 (0.103) &       &       &       &         \\
			\cmidrule{1-7}
			\multirow{2}[0]{*}{Fr\'echet} &$\hat{a}$& 0.385 (0.014) & \multirow{2}[0]{*}{2} & \multirow{2}[0]{*}{-1927.01} & \multirow{2}[0]{*}{3858.02} & \multirow{2}[0]{*}{3865.37}  \\
			& $\hat{b}$ & 209.940 (33.806) &       &       &       &         \\
			\cmidrule{1-7}
			\multirow{4}[0]{*}{DPLN} & $\hat{\lambda_1}$& 3.803 (2.759) & \multirow{4}[0]{*}{4} & \multirow{4}[0]{*}{-1874.11} & \multirow{4}[0]{*}{3756.23} & \multirow{4}[0]{*}{3770.92} \\
			& $\hat{\lambda_2}$ & 2.144 0.021) &       &       &       &        \\
			& $\hat{\tau}$ & 2.142 (0.021) &       &       &       &        \\
			& $\hat{\nu}$& 4.409 (0.022) &       &       &       &         \\
\bottomrule
		\end{tabular*}%
		\label{tab:estimates - data I}%
		\begin{tablenotes}
			\item \small *The standard errors of estimates are reported in parentheses.
		\end{tablenotes}
	\end{table}%

	\begin{table}[hbt!]
		\centering
		\caption{Earthquake economic losses: goodness-of-fit  measures.}
		\begin{tabular*}{\hsize}{@{}@{\extracolsep{\fill}}llllllll@{}}
\toprule
			\multirow{2}[0]{*}{Distribution} & \multirow{2}[0]{*}{R} &\multicolumn{2}{l}{Kolmogorov-Smirnov} & \multicolumn{2}{l}{Anderson-Darling} & \multicolumn{2}{l}{Cramer-von Mises} \\
			\cmidrule{3-8}
			&  & Statistic & P-value & Statistic & P-value & Statistic & P-value \\
			\hline
			GlogM & 0.975& 0.089 & 0.000 & 5.765 & 0.000 & 0.907 & 0.000 \\
			GLMGA &\textbf{0.997} & \textbf{0.028} & \textbf{0.810} & \textbf{0.282} & \textbf{0.589} & \textbf{0.035} & \textbf{0.710}\\
			Log-gamma &0.990 & 0.077 & 0.000 & 1.954 & 0.000 & 0.346  & 0.000 \\
			Fr\'echet &0.950 & 0.115 & 0.000 & 9.200 & 0.000 & 1.398 & 0.000 \\
			DPLN &0.991 & 0.040 & 0.302 & 0.525& 0.130 & 0.074 & 0.200 \\
\bottomrule
		\end{tabular*}%
		\label{tab:gof}%
		\begin{tablenotes}
			\item \small *The bootstrap P-values are computed using parametric bootstrap with 1000 simulation runs.
		\end{tablenotes}
	\end{table}%

	\begin{table}[tp!]
		\centering
		\caption{Earthquake economic losses: estimates of $\text{VaR}_{0.95}$, $\text{VaR}_{0.99}$ and $\text{VaR}_{0.998}$, relative difference (in percentage) with respect to the empirical VaR. }
		\begin{tabular*}{\hsize}{@{}@{\extracolsep{\fill}}ccccccc@{}}
\toprule
			Model  & 95\%  & Diff. \%  & 99\%  & Diff. \% & 99.8\%  & Diff. \% \\
			\hline
			Empirical & 2574.97 &       & 25349.91 &       & 49635.53 &  \\
			GlogM & 41844.12 & 15.25 & 4140052.88 & 162.32 & 408911187.73 & 892.48 \\
			GLMGA & 3577.63 & 0.39  & 50153.12 & 0.98  & 701400.43 & \textbf{0.53} \\
			Log-gamma & 5582.17 & 1.17  & 82065.15 & 2.24  & 1013583.58 & 1.22 \\
			Fr\'echet & 3621.54 & 0.41  & 11055.97 & -0.56 & 24068.97 & -0.95 \\
			DPLN  & 2509.55 & -0.03 & 11198.31 & -0.56 & 37585.05 & -0.92 \\	
\bottomrule
		\end{tabular*}%
		\label{tab:VaR-competing models}%
	\end{table}%

%
	
		\subsection{The numbers of casualties: marginal modelling}\label{App: casualties}
	
	The ML estimates for truncated count distribution is performed via the \texttt{gamlss} function of the \texttt{gamlss} and \texttt{gamlss.tr} package in R, and the estimates for GP distribution is performed via the \texttt{fevd} function of the \texttt{extRemes} package in R (see \cite{Gilleland2016} for details).
	
\begin{figure}[htbp]
		\centering
		\includegraphics[scale=0.45]{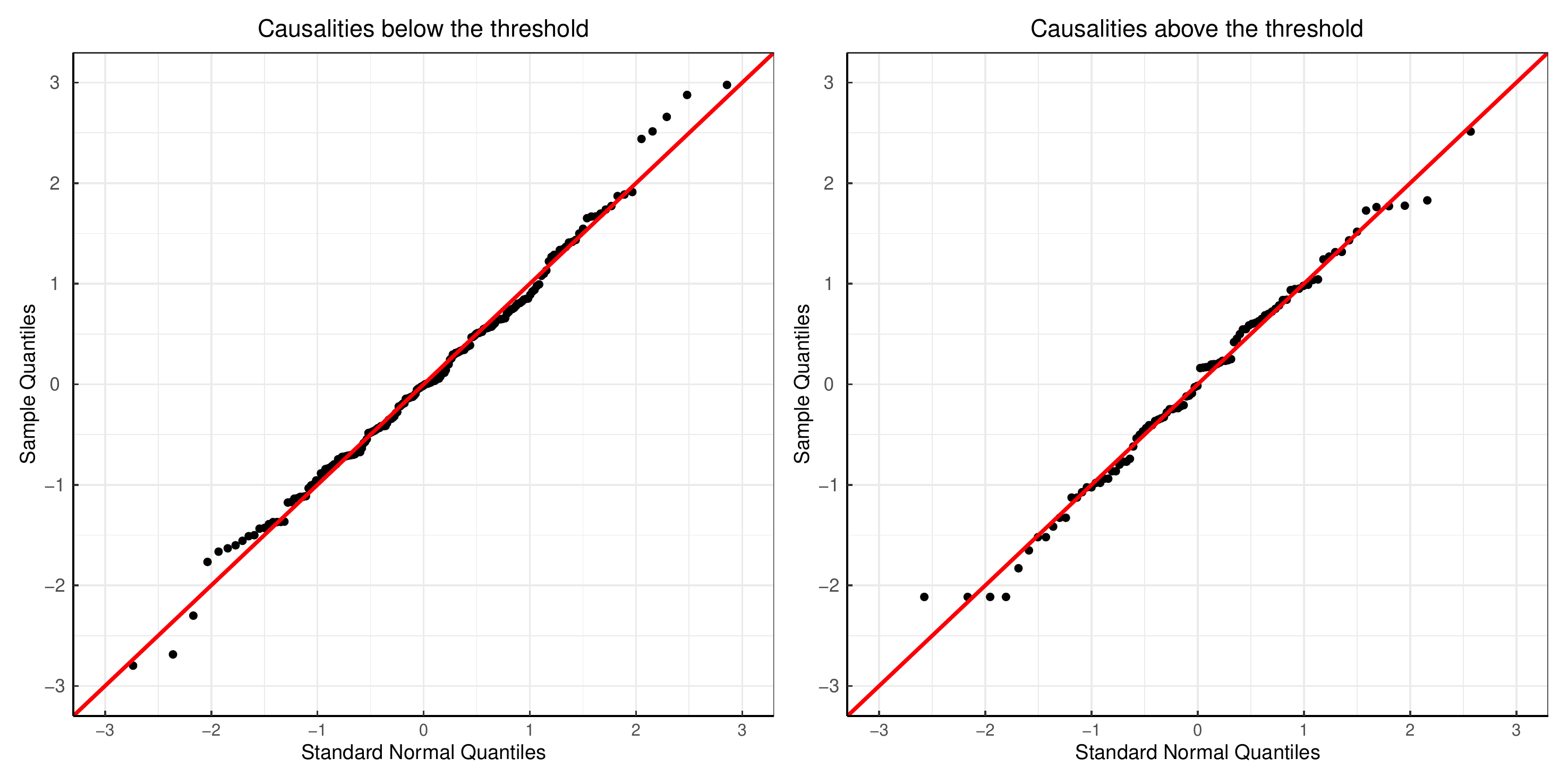} 
		\caption{
			Normal QQ-plots of quantile residuals for the number of casualties based on
			the truncated negative binomial distribution ($r_{i}^{d}$ in the \textit{left panel}), and GP distribution ($r_{i}^c$ in the \textit{right panel}) with the sample size $n-n_c$ and $n_c$ respectively.}
		\label{fig:qq-casualties}
	\end{figure}

		To demonstrate the goodness fit of the truncated count distribution below the threshold and the tail behavior above the threshold of casualties data, we use randomized (normal) quantile residuals defined by
	$r_{i}^{j}=\Phi^{-1}\left[F_{Y_2}^{j}(y_{i})\right]$
	for $i=1,\ldots,n_c$ with $j=c$ and $i=1,\ldots,n-n_c$ with $j=d$,
	where $\Phi^{-1}\left(\cdot\right)$ is the inverse function of the cdf of the standard normal distribution and $F_{Y_{j}}^{j}(\cdot)$ denotes the cdf of the right-truncated count distribution and GP distribution
	as given in \eqref{eq:F2} respectively.
	The distribution of $r_{i}^{c}$ and $r_{i}^{c}$ converge to standard normal if parameters are consistently estimated, see \cite{dunn1996randomized}, and hence a normal QQ-plot of randomized quantile residuals should follow the 45 degree line.
	Figure \ref{fig:qq-casualties} displays the normal QQ-plot for the number of casualties
	supporting the condition that residuals of right truncated negative binomial distribution and GP distribution are normally distributed.
	
	\bibliography{mybibfile}
\end{document}